\documentclass[11pt,a4paper]{article}
\pdfoutput=1 

\usepackage{jheppub} 

\usepackage[T1]{fontenc} 
\usepackage{slashed}
\usepackage[normalem]{ulem}
\usepackage{xcolor}
\usepackage{tikz}
\usepackage{mathtools}
\usetikzlibrary{decorations.pathmorphing}
\tikzset{snake it/.style={decorate, decoration=snake}}

\newcommand{\ee}[2]{\varepsilon_{#1} {\cdot} \varepsilon_{#2}}
\newcommand{\z}{x}
\newcommand{\ez}[2]{\varepsilon_{#1} {\cdot} \z_{#2}}
\newcommand{\scalarF}{\textcolor{black}{X}}

\def\eqn#1{eq.~(\ref{#1})}

\title{Lagrangians Manifesting Color-Kinematics Duality in the NMHV Sector of Yang-Mills}

\author[a]{Maor Ben-Shahar,}
\author[a]{Lucia Garozzo}
\author[a,b]{and Henrik Johansson}

\affiliation[a]{Department of Physics and Astronomy, Uppsala University, \\ Box 516, 75120 Uppsala, Sweden}
\affiliation[b]{Nordita, Stockholm University and KTH Royal Institute of Technology, \\ Hannes Alfv\'{e}ns v\"{a}g 12, 10691 Stockholm, Sweden}

\emailAdd{benshahar.maor@physics.uu.se}
\emailAdd{garozzo.lucia@physics.uu.se}
\emailAdd{henrik.johansson@physics.uu.se}

\abstract{Scattering amplitudes in Yang-Mills theory are known to exhibit kinematic structures which hint to an underlying kinematic algebra that is dual to the gauge group color algebra. This color-kinematics duality is still poorly understood in terms of conventional Feynman rules, or from a Lagrangian formalism. In this work, we present explicit Lagrangians whose Feynman rules generate duality-satisfying tree-level BCJ numerators, to any multiplicity in the next-to-MHV sector of pure Yang-Mills theory. Our Lagrangians make use of at most three pairs of auxiliary fields (2,1,0-forms) -- surprisingly few compared to previous attempts of Lagrangians at low multiplicities. To restrict the Lagrangian freedom it is necessary to make several non-trivial assumptions regarding field content, kinetic terms, and interactions, which we discuss in some detail. Future progress likely hinges on relaxing these assumptions.}

\preprint{UUITP-62/22 
\\ \phantom{~} \hfill NORDITA 2022-155
}  

\begin{document} 
\maketitle
\flushbottom

\section{Introduction}

Scattering amplitudes offer valuable insight to the mathematical structures of quantum field theory and gravity by uncovering patterns that are not readily apparent through standard Lagrangian techniques. 
The Bern-Carrasco-Johansson (BCJ) duality between color and kinematics~\cite{Bern:2008qj,Bern:2010ue,Bern:2019prr} is a clear example where on-shell formulations preceded any Lagrangian understanding. According to the duality, scattering amplitudes in many gauge theories can be represented using cubic diagrams, each consisting of a kinematic numerator and a color factor that obey isomorphic relations. The color factors draw their characteristics from the Lie algebra of the gauge group, whereas the mathematical structure of the numerators is believed to originate from an unknown kinematic Lie algebra. See recent reviews~\cite{Bern:2019prr,Bern:2022wqg,Adamo:2022dcm,McLoughlin:2022ljp,Berkovits:2022ivl,Bern:2022jnl,Mafra:2022wml}. 

In massless purely adjoint gauge theories, such as pure Yang-Mills (YM) theory, color-kinematics duality can be rephrased at tree level as the presence of BCJ amplitude relations~\cite{Bern:2008qj,Stieberger:2009hq,BjerrumBohr:2009rd,Feng:2010my,BjerrumBohr:2010hn}. The duality and amplitude relations were initially identified in pure YM~\cite{Bern:2008qj} and its supersymmetric generalizations~\cite{Bern:2010ue,Bern:2010yg,Stieberger:2009hq,BjerrumBohr:2009rd}, but have since been discovered in a range of other gauge theories, including matter representations~\cite{Chiodaroli:2013upa,Johansson:2014zca,Chiodaroli:2014xia,Johansson:2015oia,Chiodaroli:2015rdg,Chiodaroli:2018dbu,Johansson:2019dnu,Bautista:2019evw,Plefka:2019wyg}, higher-derivative interactions~\cite{Broedel:2012rc,Johansson:2017srf,Johansson:2018ues,Azevedo:2018dgo, Garozzo:2018uzj,Carrasco:2021ptp,Chi:2021mio,Menezes:2021dyp, Bonnefoy:2021qgu,Carrasco:2022lbm,Carrasco:2022jxn,Carrasco:2022sck} or Chern-Simons fields~\cite{Bargheer:2012gv,Huang:2012wr,Huang:2013kca, Sivaramakrishnan:2014bpa,Ben-Shahar:2021zww}, and massive gauge theories~\cite{Naculich:2014naa,Naculich:2015zha,Johansson:2015oia,Chiodaroli:2015rdg,Johansson:2019dnu,Momeni:2020vvr, Johnson:2020pny, Moynihan:2020ejh,Momeni:2020hmc,Gonzalez:2021bes,Moynihan:2021rwh, Gonzalez:2021ztm,Chiodaroli:2022ssi,Li:2021yfk,Gonzalez:2022mpa,Li:2022rel,Emond:2022uaf,Engelbrecht:2022aao}.
Surprisingly, also certain scalar effective field theories~\cite{Chen:2013fya,Cheung:2016prv,Carrasco:2016ldy,Mafra:2016mcc,Carrasco:2016ygv,Low:2019wuv,Cheung:2020qxc,Rodina:2021isd,deNeeling:2022tsu, Cheung:2022vnd} obey the duality. The duality has been extended to loop-level amplitudes~\cite{Bern:2010ue, Carrasco:2011mn, Bern:2012uf, Boels:2013bi, Bjerrum-Bohr:2013iza, Bern:2013yya, Nohle:2013bfa, Mogull:2015adi, Mafra:2015mja, He:2015wgf,Johansson:2017bfl, Hohenegger:2017kqy, Mafra:2017ioj, Faller:2018vdz, Kalin:2018thp, Ben-Shahar:2018uie, Duhr:2019ywc, Geyer:2019hnn, Edison:2020uzf, Casali:2020knc, DHoker:2020prr, Carrasco:2020ywq, Bridges:2021ebs,Guillen:2021mwp,Porkert:2022efy,Edison:2022smn,Edison:2022jln}, form factors~\cite{Boels:2012ew,Yang:2016ear,Boels:2017ftb,Lin:2020dyj,Lin:2021kht,Lin:2021pne,Lin:2021lqo, Lin:2021qol,Chen:2022nei,Li:2022tir}, and even to curved-space correlators~\cite{Adamo:2017nia,Farrow:2018yni,Adamo:2018mpq,Lipstein:2019mpu, Prabhu:2020avf, Armstrong:2020woi,Albayrak:2020fyp,Adamo:2020qru,Alday:2021odx,Diwakar:2021juk,Drummond:2022dxd,Herderschee:2022ntr,Zhou:2021gnu,Sivaramakrishnan:2021srm,Alday:2022lkk,Cheung:2022pdk,Bissi:2022wuh,Li:2022tby,Lee:2022fgr}. There have been various approaches to understanding the existence of color-kinematics duality and BCJ relations in massless gauge theories, including string theory, scattering equations, and positive geometry~\cite{BjerrumBohr:2009rd,Stieberger:2009hq,Cachazo:2012uq,Arkani-Hamed:2017mur,Mizera:2019blq,Britto:2021prf,Ahmadiniaz:2021fey,Ahmadiniaz:2021ayd}.

Color-kinematics duality is intimately connected to the double-copy structure of gravity~\cite{Bern:2008qj,Bern:2010ue}. For non-abelian gauge theories that obey color-kinematics duality, and contain physical spin-1 gluons, the double copy gives gravitational interactions after taking each cubic diagram and replacing the color factor by a second numerator copy. The double copy provides a versatile generalization of the Kawai-Lewellen-Tye (KLT) relations~\cite{Kawai:1985xq}, which were originally derived for string theory at genus zero (for higher-genus generalizations see e.g.~\cite{DHoker:1989cxq,Geyer:2015jch,Geyer:2016wjx,He:2016mzd,He:2017spx,Geyer:2019hnn,Casali:2020knc,Geyer:2021oox,Stieberger:2022lss}). Through color-kinematics duality the double copy permits the construction of gravity loop amplitudes~\cite{Bern:2010ue,Bern:2011rj, BoucherVeronneau:2011qv, Bern:2013uka,Bern:2014sna,Chiodaroli:2015wal,Johansson:2017bfl,Chiodaroli:2017ehv,Bern:2018jmv,Bern:2021ppb}, more general gravity theories~\cite{Broedel:2012rc, Chiodaroli:2013upa,Johansson:2014zca,Chiodaroli:2014xia,Johansson:2015oia,Chiodaroli:2015rdg,Johansson:2017srf,Chiodaroli:2018dbu,Johansson:2018ues,Azevedo:2018dgo,Johansson:2019dnu,Bautista:2019evw,Plefka:2019wyg,Pavao:2022kog, Mazloumi:2022nvi}, classical solutions~\cite{Monteiro:2014cda,Luna:2015paa,Luna:2016hge,Bahjat-Abbas:2017htu,Carrillo-Gonzalez:2017iyj,Berman:2018hwd,CarrilloGonzalez:2019gof,Goldberger:2019xef,Huang:2019cja,Bahjat-Abbas:2020cyb,Easson:2020esh,Emond:2020lwi, Godazgar:2020zbv,Chacon:2021wbr,Chacon:2020fmr,Alfonsi:2020lub, Monteiro:2020plf, White:2020sfn, Elor:2020nqe,Pasarin:2020qoa, Adamo:2021dfg, Easson:2022zoh,Dempsey:2022sls,CarrilloGonzalez:2022ggn} and other non-perturbative solutions~\cite{Cheung:2022mix,Armstrong-Williams:2022apo}. Recently black-hole scattering and gravitational wave physics have been studied using the double copy~\cite{Luna:2016due,Goldberger:2016iau,Luna:2017dtq,Shen:2018ebu,Plefka:2018dpa,Bern:2019nnu,Plefka:2019hmz,Bern:2019crd,Bern:2020buy,Almeida:2020mrg,Haddad:2020tvs,Bern:2021dqo,Bern:2021yeh,Bern:2022kto, Carrasco:2020ywq,
Carrasco:2021bmu,Chiodaroli:2021eug,Shi:2021qsb,CarrilloGonzalez:2022mxx,Cangemi:2022abk,Bjerrum-Bohr:2022ows,Comberiati:2022cpm}.

Despite extensive progress, the mathematical details of color-kinematics duality remain enigmatic even for the case of pure YM theory. While duality-satisfying YM numerators can be straightforwardly computed to any multiplicity at tree level~\cite{BjerrumBohr:2010hn,Mafra:2011kj,Mafra:2015vca,Bjerrum-Bohr:2016axv,Du:2017kpo,Chen:2017bug,Fu:2018hpu,Edison:2020ehu,Hou:2021mvg,Cheung:2021zvb,Brandhuber:2021bsf,Brandhuber:2022enp,Cao:2022vou}, the building blocks and underlying symmetry principles of these objects are poorly understood. Realizing the hidden YM structure as a kinematic Lie algebra or a cubic Lagrangian may shine light on the duality. Explicit kinematic-algebra constructions have been realized for the self-dual YM sector~\cite{Monteiro:2011pc,Boels:2013bi,Monteiro:2022lwm}, the maximally-helicity-violating (MHV) sector~\cite{Monteiro:2011pc,Cheung:2016prv,Chen:2019ywi} and the next-to-MHV (NMHV) sector~\cite{Chen:2019ywi,Chen:2021chy}. 
Self-dual YM amplitudes vanish except for at one loop~\cite{Cangemi:1996rx} and the kinematic algebra is identified with area-preserving diffeomorphisms~\cite{Monteiro:2011pc}. 
The MHV sector of YM gives the simplest non-vanishing 4D tree amplitudes, and while ref.~\cite{Monteiro:2011pc} considered it through a non-local gauge choice, local formulations of the MHV kinematic algebra was discussed in refs.~\cite{Cheung:2016prv,Chen:2019ywi}. 

A cubic duality-satisfying Lagrangian for the  non-linear-sigma model was considered in ref.~\cite{Cheung:2016prv}, but via dimensional-reduction operations~\cite{Cheung:2017yef,Cheung:2017ems} it is directly related to the MVH sector of YM. (It is also related to a non-abelian generalization of the Navier-Stokes equation that obeys color-kinematics duality~\cite{Cheung:2020djz,Keeler:2020rcv,Escudero:2022zdz}.) Both the MHV and NMHV kinematic algebra was described in some detail in refs.~\cite{Chen:2019ywi,Chen:2021chy}. A natural extension of the area-preserving diffeomorphism algebra remains to be found in those sectors. However, in 3D Chern-Simons theory the complete kinematic algebra is now identified with volume-preserving diffeomorphisms~\cite{Ben-Shahar:2021zww}. In section~\ref{MHVsection}, we briefly comment on the appearance of 4D volume-preserving diffeomorphism generators in the MHV sector. Recently, kinematic Hopf algebras have been introduced for YM and related theories, which gives gauge-invariant BCJ numerators for all tree-level sectors at the expense of additional heavy-mass poles~\cite{Cheung:2021zvb,Brandhuber:2021kpo,Brandhuber:2021eyq,Brandhuber:2021bsf,Chen:2022nei,Brandhuber:2022enp,Cao:2022vou}.

Duality-satisfying YM Lagrangians have been sporadically studied for more than a decade~\cite{Bern:2010yg, Tolotti:2013caa, Ferrero:2020vww,Borsten:2020zgj, Borsten:2020xbt, Beneke:2021ilf, Lam:2021vly,Ben-Shahar:2021doh} using non-local interaction terms or auxiliary fields. The early attempts produced Feynman rules capable of computing local BCJ numerators up to five~\cite{Bern:2010yg} and six points~\cite{Tolotti:2013caa}. However, the non-uniqueness of the BCJ numerators led to a proliferation of ambiguities in the resulting Lagrangians. In hindsight, it is clear that in a local formalism the ambiguities are absent in the MHV sector and first appear in the NMHV sector~\cite{Chen:2019ywi}. They can be traced back to the so-called generalized gauge freedom of the BCJ numerators~\cite{Bern:2008qj,Bern:2010ue}, which include the standard gauge freedom and non-linear field redefinitions. Understanding how to constrain this freedom and write down broadly valid and practical YM Lagrangians that manifest the duality is the topic of this paper.  

Recently there has been an upswing in formulations that attempt to approach the problem of realizing YM color-kinematics duality and double copy off shell~\cite{Anastasiou:2018rdx,Bridges:2019siz,Borsten:2020xbt,Borsten:2020zgj,  Cheung:2021zvb, Campiglia:2021srh, Borsten:2021zir, Cho:2021nim, Borsten:2021rmh, Bonezzi:2022bse, Diaz-Jaramillo:2021wtl,Ben-Shahar:2021zww,Ben-Shahar:2021doh,Godazgar:2022gfw,Bonezzi:2022yuh,Cheung:2022mix,Borsten:2022ouu,Borsten:2022vtg}. This involves understanding the role of BRST symmetries, homotopy algebras, twistor spaces, double field theory, equations of motions and other off-shell Lagrangian perspectives. We note that, in principle, the problem has been addressed using the 10D pure-spinor formalism~\cite{Ben-Shahar:2021doh}; however, the practical usefulness of this duality-satisfying supersymmetric YM Lagrangian for advanced calculations remains to be understood.

In this paper, we approach the problem of finding duality-satisfying YM Lagrangians head on, by making a suitable small ansatz for the  NMHV sector Lagrangian where the first ambiguities show up. We work in general dimension using covariant building blocks, and thus we generalize the 4D notion of helicity sectors by grading the tree-level numerators by the ``polarization power''~\cite{Chen:2019ywi,Chen:2021chy}, i.e.~the number of inner products between polarization vectors $\varepsilon_i {\cdot} \varepsilon_j$. Thus the NMHV sector corresponds to numerators with at most two polarization powers $(\varepsilon_i {\cdot} \varepsilon_j)^2$; or, equivalently, one power of momentum inner products $p_i {\cdot} p_j$. The construction of the ansatz is aided by the observation that there exist a bi-scalar subsector\footnote{Terms in the half-ladder BCJ numerator proportional to a fixed polarization product $\varepsilon_1 {\cdot} \varepsilon_n$, which effectively becomes two scalars interacting with the YM field.} of the NMHV sector that has particularly simple BCJ numerators. This bi-scalar subsector was also considered in refs.~\cite{Chiodaroli:2017ngp,Chen:2019ywi} since it generates master numerators from which all other NMHV numerators can be determined. We find that the bi-scalar numerators can be computed using an exact rewriting of the standard YM tree-level Lagrangian, by ``integrating in'' a pair of auxiliary 2-form fields $\{B^{\mu\nu},\tilde B^{\mu\nu}\}$ with additional simple cubic interactions.    

From the bi-scalar numerators one can obtain the remaining contributions to the NMHV numerators using a simple formula; however, we also demanded that those contributions come directly from a Lagrangian. This required us to either use clever inspection and identification of needed new interaction terms aided by pictorial diagrams that expose the tensors structures of intermediate fields, or alternatively, a brute-force Lagrangian ansatz where the assumptions are more clearly spelled out. We find several interesting solutions which involve a pair of auxiliary vector fields $\{Z^{\mu},\tilde Z^{\mu}\}$, and a pair of auxiliary scalar fields $\{X,\tilde X\}$. We show that by considering an extended ansatz the need for scalar auxiliary fields is not obvious, as there exist interesting Lagrangian solutions where they are absent. Further relaxing the assumptions that went into the construction may provide simpler NMHV Lagrangians, but we leave this for future work. We briefly comment on the N${}^2$MHV sector, and also spell out some details about the possible generalization of the Lagrangian Feynman rules to one-loop calculations.

The paper is organized as follows: In section~\ref{sec:preliminaries}, we introduce the necessary notation used for describing amplitudes and numerators that satisfy color-kinematics duality, including decomposition into partial amplitudes, bi-scalar half-ladder numerators and useful pictorial diagrams. In section~\ref{MHVsection}, we consider a cubic Lagrangian for computing MHV numerators, obtained by truncating the standard YM action. In section~\ref{NMHVsection}, we transform the standard YM Lagrangian by integrating in a pair of two-form auxiliary fields and then we add interactions and further auxiliary fields using both diagrammatic and ansatz approaches.  In section~\ref{sec:one-loop}, we  briefly discuss consequences for one-loop numerators.  Conclusions and outlook are given in section~\ref{sec:conclucions}.

\section{Graphs, Numerators and Tree Amplitudes \label{sec:preliminaries}}

Here we will set the notation used for the amplitude and numerator building blocks, and discuss the decompositions and diagrammatic notation that are convenient for later sections.  

\subsection{Color-kinematics duality and double copy}

Scattering amplitudes in YM theory can be represented diagrammatically as a sum over cubic Feynman-like graphs~\cite{Bern:2008qj,Bern:2010ue}, for $n$-point tree level amplitudes this takes the form\footnote{Overall imaginary factors are suppressed throughout the paper.}
\begin{equation}
\label{NumAmplFormula}
\mathcal{A}_n= g^{n-2}\sum_{\Gamma\in{\cal G}_{n}}\frac{C_\Gamma N_\Gamma}{D_\Gamma} \, .
\end{equation}
Each cubic graph  $\Gamma$ is associated with a color factor $C_\Gamma$, kinematic numerator $N_\Gamma$ and propagator denominator $D_\Gamma$. 
Here the set of cubic $n$-point graphs is denoted by ${\cal{G}}_n$, and it can be constructed recursively starting from the three-point case, ${\cal{G}}_3$, which contains only one such graph $\Gamma=[1,2]$. At $n$ points the set of cubic graphs is given by
\begin{equation}
{\cal G}_n = \Big\{\Gamma|_{\ell \rightarrow [\ell,\, n-1]} ~\Big|~ \ell \in \Gamma \in  {\cal G}_{n-1} \Big\}\,,
\end{equation}
where $\ell$ denotes a Lie-valued element of the graph $\Gamma$. For example, the graph $\Gamma=[1,2]$ has three such elements, the external leg labels $\ell=1$, $\ell=2$ and the commutator of the labels $\ell=[1,2]$. Applying the rule $\ell \rightarrow [\ell,3]$ then gives three four-point graphs represented by nested commutators~\cite{Mafra:2020qst,Frost:2020eoa, Chen:2021chy}
\begin{equation}
{\cal G}_4 = \Big\{ [[1,3],2],~  [1,[2,3]],~  [[1,2],3] \Big\}\,.
\end{equation}
Each one of the four-point graphs have five Lie-valued elements (three labels, and two commutators), thus the total number of five-point graphs is $|{\cal G}_5|= 3\times 5$. In general, at $n$ points, the number of cubic graphs is  $|{\cal G}_n|= 3\times 5 \times 7 \times \cdots \times (2n-5)$ = $(2n-5)!!$.   

With the above graph notation the propagator denominator can be written as 
\begin{equation} \label{prop_denominator}
D_{\Gamma}=\!\!\!\!\prod_{\substack{\ell \in  \Gamma \\ \ell\neq \mathbb{N}, \ell\neq\Gamma}} \!\! p_\ell^2\,,
\end{equation}
where $p_\ell$ is the sum of the momentum of all the external legs in the nested commutator $\ell$. For example, $p_{[1,2]}=p_1+p_2$ and  $p_{[[1,3],2]}=p_1+p_2+p_3$, etc. Since the external legs are on shell $p_1^2,\ldots, p_n^2=0$, we do not include cases where $\ell$ is an integer, nor where it is the graph $\Gamma$, since by momentum conservation $p_\Gamma= -p_n$.

The color factors for purely-adjoint YM theory correspond to rank-$n$ tensors in the gauge-group Lie algebra, which can be constructed through contraction of structure constants, $f^{abc}$, or as traces of products of generators $T^a$. The latter case is most transparent in our graph notation, and an explicit formula can be given
\begin{equation}
C_{\Gamma}= C_{\Gamma}^{a_1 a_2 \cdots a_n}= {\rm Tr} \big\{(\Gamma |_{i\rightarrow T^{a_i}})  \, T^{a_n}  \big\} \,,
\end{equation}
where the external legs $i=1,\ldots n-1$ of the graph $\Gamma$ are replaced by Lie-algebra generators, thus justifying the graph notation as nested commutators of Lie-valued elements. For example, the three-point graph $\Gamma=[1,2]$ has the color factor
\begin{equation}
C_{[1,2]} = {\rm Tr} \big\{[T^{a_1},T^{a_2}] T^{a_3}\big\} = f^{a_1 a_2 a_3}\,.
\end{equation}
Here, and throughout the paper, we use the normalization conventions $[T^{a},T^{b}]= f^{abc} T^{c}$ and ${\rm Tr} \big\{ T^{a} T^{b}\big\}= \delta^{ab}$.

Finally the numerators $N_{\Gamma}$ capture all the remaining dependence on the kinematic data of the YM tree amplitude, and in this paper we assume that they are {\it local} polynomials of the polarization vectors $\varepsilon_i$ and momenta $p_i$. To describe gluon scattering in pure YM theory, the polynomials must be multi-linear in the individual polarizations and of degree-$(n-2)$ in the momenta. It implies that the contributing monomial terms have the schematic form 
\begin{equation} \label{NumeratorStructure}
N_{\Gamma}  ~ \sim ~ \sum_{k=1}^{\lfloor n/2 \rfloor} \,(\varepsilon  \cdot\varepsilon)^{k}\,(p \cdot p )^{k-1} \, (\varepsilon \cdot p )^{n-2k}\,.
\end{equation}
For example, for $n=3$ points the numerator is equivalent to the color-stripped cubic Feynman rule of YM,
\begin{equation}
N_{[1,2]} = \varepsilon_1  \cdot\varepsilon_2\,  \varepsilon_3  \cdot (p_2-p_1) + {\rm cyclic}(1,2,3)\,.
\end{equation}
It satisfies the same dihedral permutation symmetries $N_{[2,1]}=-N_{[1,2]}$, $N_{[2,3]}=N_{[3,1]}=N_{[1,2]}$ as the color factor $C_{[1,2]}=f^{a_1 a_2 a_3}$. 

Due to the Jacobi identity of the gauge-group Lie algebra, there are certain cubic relations between color factors. The statement of the color-kinematics duality~\cite{Bern:2008qj, Bern:2010ue} is that for a large class of gauge theories there exists a choice of so-called BCJ numerators that obey the same relations as the color factors,
\begin{align} \label{JacobiId}
    &C_{\cdots[[X,Y],Z]\cdots} + C_{\cdots[[Y,Z],X]\cdots}+ C_{\cdots[[Z,X],Y]\cdots} = 0 \nonumber \\ & \hskip 3.5cm \Leftrightarrow \notag \\
    &N_{\cdots[[X,Y],Z]\cdots} +  N_{\cdots[[Y,Z],X]\cdots}+ N_{\cdots[[Z,X],Y]\cdots} = 0 \, .
\end{align}
A four-point YM numerator that satisfies this property is~\cite{Bern:2019prr}
\begin{eqnarray}\label{four_point_numerator}
N_{[[1,2],3]} &=&
\big(\varepsilon_1{\cdot}\varepsilon_2 p_1^\mu + 2 \varepsilon_1{\cdot}p_2 \varepsilon_2^\mu -(1\leftrightarrow 2)\big)
\big(\varepsilon_3{\cdot}\varepsilon_4 p_{3\mu} + 2 \varepsilon_3{\cdot}p_4 \varepsilon_{4\mu} -(3\leftrightarrow 4)\big)\nonumber\\
&& \null +s_{12} (\varepsilon_1{\cdot} \varepsilon_3 \varepsilon_2{\cdot} \varepsilon_4-\varepsilon_1{\cdot} \varepsilon_4 \varepsilon_2{\cdot} \varepsilon_3)\,.
\end{eqnarray}
It is straightforward to check that, subject to on-shell conditions and momentum conservation, the numerator satisfies the Jacobi relation
\begin{equation}
N_{[[1,2],3]}+N_{[[2,3],1]}+N_{[[3,1],2]}=0\,,
\end{equation}
as well as the permutation symmetries
\begin{equation}
N_{[[2,1],3]}=-N_{[[1,2],3]}\,,~~~~ N_{[[4,3],2]}=N_{[[1,2],3]}\,.
\end{equation}
In general, it is a difficult task to find YM numerators that satisfy color-kinematics duality. However, efforts during the last decade have led to several different constructions of all-multiplicity numerators with various properties~\cite{BjerrumBohr:2010hn,Mafra:2011kj,Mafra:2015vca,Bjerrum-Bohr:2016axv,Du:2017kpo,Chen:2017bug,Edison:2020ehu,Bjerrum-Bohr:2020syg, Hou:2021mvg,Cheung:2021zvb,Brandhuber:2021bsf, Ahmadiniaz:2021ayd,Brandhuber:2022enp}. Nevertheless, until now it has not been known how to construct such numerators directly from a duality-satisfying Lagrangian or Feynman rules. We will attempt to address this problem in section~\ref{NMHVsection}.

The color factors in \eqn{NumAmplFormula} can be substituted by a second copy of color-kinematics satisfying numerators $C_\Gamma \rightarrow \tilde N_\Gamma$, which gives the double-copy formula~\cite{Bern:2008qj, Bern:2010ue}
\begin{equation}
    \mathcal{M}_n = \left(\frac{\kappa}{2}\right)^{n-2}\sum_{\Gamma\in{\cal G}_n}\frac{N_\Gamma\tilde N_\Gamma}{D_\Gamma} \,,
\end{equation}
where, for dimensional consistency, we also substituted the coupling $g\rightarrow \kappa/2$. The two sets of numerators can belong to either the same or different theories. 
If both sets of numerators belong to theories that contain physical spin-1 gauge fields, then the double-copy formula gives an amplitude for spin-2 fields in some theory of gravity. This follows from the fact that the double copy automatically gives diffeomorphism invariant amplitudes. To see this, consider a linearized gauge transformation acting on one of the polarizations of the gauge theory amplitude, $\delta \varepsilon_i = p_i$.  By definition, we expect that  $\delta {\cal A}_n = 0$, from which it follows that it must be true that  
\begin{equation}
 \sum_{\Gamma\in{\cal G}_n}\frac{C_\Gamma\, \delta N_\Gamma}{D_\Gamma}=0\,,
\end{equation}
where $\delta N_\Gamma$ are the gauge transformed numerators. The fact that this vanishes does not depend on the details of the color factors, only on the fact that they come from a Lie algebra and thus satisfy Jacobi identities. If we have kinematic numerators that satisfy the same Jacobi identities, then it also follows that~\cite{Bern:2008qj, Bern:2010ue,Chiodaroli:2017ngp}  
\begin{equation}
 \sum_{\Gamma\in{\cal G}_n}\frac{\tilde N_\Gamma\, \delta N_\Gamma}{D_\Gamma}=0\,,
\end{equation}
and likewise for $N_\Gamma$ and $\tilde N_\Gamma$ swapped. Hence, it follows that $\delta {\cal M}_n = 0$, and thus the double-copy formula is invariant under linearized diffeomorphisms~\cite{Chiodaroli:2017ngp}. Given that all amplitudes have this invariance, it follows that the theory enjoys non-linear diffeomorphism symmetry, so it can be interpreted as a theory of gravity. When both numerators come from four-dimensional pure YM, the double copy gives amplitudes in Einstein-dilaton-axion gravity, with Lagrangian~\cite{Bern:2019prr}
\begin{equation}
{\cal L} =-\frac{\sqrt{-g}}{\kappa^2}\Big(2R -\frac{\partial_\mu \tau \partial^\mu \bar \tau}{({\rm Im}\, \tau)^2}\Big)\,,
\end{equation}
where the complex field $\tau=ie^{-\phi} +\chi$ contains the dilaton $\phi$ and axion $\chi$ scalars. In $D$ dimensions the axion is promoted to a two-form $B^{\mu \nu}$ and the Lagrangian is also well known, see e.g.~\cite{Bern:2019prr}. 

\subsection{Decompositions for amplitudes and numerators}

Color-ordered partial amplitudes are defined as the gauge-invariant coefficients of the trace decomposition of the full amplitude,
\begin{equation}
    \mathcal{A}_n = g^{n-2}\sum_{\sigma\in S_{n-1}} \textrm{Tr}(T^{a_1}T^{a_{\sigma(2)}}\ldots T^{a_{\sigma(n)}})A_n(1,\sigma_{2},\ldots,\sigma_n) \, ,
\end{equation}
where the sum over permutations $\sigma$ runs over the symmetric group $S_{n-1}$ with $(n-1)!$ elements.
The partial amplitudes can be computed directly using color-ordered Feynman rules, or from \eqn{NumAmplFormula} after expanding out the color factors $C_\Gamma$ into the basis of ordered traces of generators. 

Alternatively, repeatedly using the Jacobi identity \eqref{JacobiId} to obtain a smaller basis of color factors, gives the Del-Duca-Dixon-Maltoni (DDM) decomposition  \cite{DelDuca:1999rs}, 
\begin{equation} 
    \mathcal{A}_n = g^{n-2}\sum_{\sigma\in S_{n-2}} C(1,\sigma_2,\ldots,\sigma_{n-1},n) 
    A_n(1,\sigma_2,\ldots,\sigma_{n-1},n) \, ,
\end{equation}
which uses a subset of the same partial amplitudes, and the independent color factors are
\begin{align}
C(1,\sigma_2,\ldots,\sigma_{n-1},n)&\equiv C_{[[\cdots [[1,\rho_2],\rho_3], \ldots ], \rho_{n-1}]} ={\rm Tr}( [[\cdots[T^{a_1}, T^{a_{{\sigma(2)}}}],\ldots], T^{a_{{\sigma(n-1)}}}] T^{a_n} ) \\ & = \big(f^{a_{\sigma(2)} } \ldots f^{a_{\sigma(n-1)}} )_{a_1 a_n}\,.
\end{align}
The $(n-2)!$-fold DDM basis of color factors $C_\Gamma$  makes use of the so-called half-ladder graphs 
\begin{equation}
\Gamma_\text{half-ladder}(\rho) \equiv {[[\cdots [[1,\rho_2],\rho_3], \ldots ], \rho_{n-1}]}\,,
\end{equation}
which will serve as our canonical basis choice of graphs. 
Similarly, repeatedly using the kinematic Jacobi identity \eqref{JacobiId} on the BCJ numerators gives the decomposition~\cite{Bern:2008qj,Vaman:2010ez}
\begin{align} \label{amplitude_in_basis}
\mathcal{A}_n=\sum_{\Gamma\in{\cal G}_n}\frac{C_\Gamma\,N_\Gamma}{D_\Gamma} = \sum_{\sigma,\rho \in S_{n-2}}  C(1,\sigma_2,\cdots,\sigma_{n-1},n) m(\sigma | \rho) N(1,\rho_2\cdots,\rho_{n-1},n)\,,
\end{align}
where we have set $g=1$ and the independent half-ladder numerators are written as
\begin{equation}
N(1,\rho_2\cdots,\rho_{n-1},n) \equiv N_{[[\cdots [[1,\rho_2],\rho_3], \ldots ], \rho_{n-1}]}\,.
\end{equation}
Assuming that the half-ladder numerators come from a manifestly crossing-symmetric construction, they obey a reflection symmetry $N(1,2,3\ldots,n) = (-1)^n N(n,\ldots,3,2,1)$, as well as anti-symmetry in the first and last pairs of arguments:  $N(2,1\ldots) =  -N(1,2\ldots)$ and $N(\ldots,n,n-1) =-  N(\ldots,n-1,n)$.

The matrix $m(\sigma | \rho)$ in \eqn{amplitude_in_basis} is of size $(n{-}2)!\times (n{-}2)!$ and is built out of linear combinations of the scalar-type propagators, $1/D_\Gamma$, as can be worked out from the above two DDM decompositions of the BCJ numerators and color factors. The matrix $m(\sigma | \rho)$ has appeared in many places in the literature, and it goes by a variety of different names, such as ``propagator matrix''~\cite{Vaman:2010ez}, the ``bi-adjoint scalar amplitude''~\cite{Cachazo:2013iea}, or ``inverse of the KLT kernel''~\cite{Kawai:1985xq,Cachazo:2013iea}. It is also equivalent to the color-stripped partial amplitudes of ``dual-scalar theory''~\cite{Bern:2010yg,BjerrumBohr:2012mg}, ``color-scalar theory''~\cite{Du:2011js} or ``scalar $\phi^3$ theory''~\cite{Bern:1999bx,Chiodaroli:2014xia}.

Using the propagator matrix the color-ordered partial amplitudes can be written as
\begin{align} \label{eq:BAampli}
A(1,\sigma,n) =& \sum_{\rho \in S_{n-2}} m(\sigma | \rho) N(1,\rho,n)\,,
\end{align} 
which gives a (non-invertible) map between BCJ numerators and partial amplitudes~\cite{Bern:2008qj}.
The propagator matrix is not invertible for on-shell conserved momenta and hence it has a kernel, or null space. This implies that BCJ numerators contain unphysical contributions that live in this kernel. This explains why BCJ numerators are in general not unique.  The ambiguity is called \emph{generalized gauge freedom}~\cite{Bern:2008qj,Bern:2010ue} and it corresponds to the freedom of shifting the BCJ numerators by \emph{pure gauge} contributions
\begin{align}\label{eq:DefinePureGauge}
N(1,\rho ,n)\sim N(1,\rho ,n)+N^{\text{gauge}}(1,\rho ,n)\,,
\end{align} 
where pure gauge numerators are annihilated by the propagator matrix,
\begin{align}
\sum_{\rho\in S_{n-2}}m(\sigma|\rho)N^{\text{gauge}}(1,\rho,n)=0\,. 
\end{align} 
The generalized gauge freedom includes both standard gauge freedom and field redefinitions, and more generally any operation that changes the cubic diagram numerators while leaving the partial amplitudes invariant. 

As illustrated in~\eqn{NumeratorStructure}, tree-level YM numerators are polynomial functions of the independent Lorentz contractions of polarizations and momenta $\{\varepsilon_i\cdot\varepsilon_j,\varepsilon_i\cdot p_j,p_i\cdot p_j\}$. Most numerator operations that we are interested in do not mix terms that contains different powers $\sim (\varepsilon_i\cdot\varepsilon_j)^k$, so it is convenient to decompose the numerators into sectors defined by their {\it polarization power} $k$~\cite{Chen:2019ywi}. Thus from the general structure~\eqref{NumeratorStructure}, we decompose the numerators as 
\begin{equation}
N(1,\ldots, n) = \sum_{k=1}^{\lfloor n/2 \rfloor} N^{(k)}(1,\ldots, n) \, ,
\end{equation}
where the polarization power $k$ keeps track of how many $\varepsilon_i {\cdot}\varepsilon_j$-contractions are present in each monomial. This decomposition also makes physical sense since by a suitable choice of the reference momenta $q_i^\mu$ in the polarizations $\varepsilon_i=\varepsilon_i(p_i,q_i)$ one can show~\cite{Chen:2019ywi} that only the numerators $N^{(1)},N^{(2)},\ldots, N^{(k)}$ contribute to the N${}^{k-1}$MHV sector of YM. Therefore when we refer to the N${}^{k-1}$MHV sector in this paper, we implicitly mean that we consider the $N^{(\le k)}$ numerators that contribute to this sector.  

Alternatively, if one considers a dimensional reduction of YM, $SO(1,D{-}1)\rightarrow SO(1,3) \times SO(D{-}4)$, then the $N^{(k)}$ numerators give rise to tree amplitudes with at most $2k$ external scalars. This can be formalized by considering derivative operators $\frac{\partial}{\partial\varepsilon_i{\cdot}\varepsilon_j}$ that can act on the numerators and convert a pair of gluons to a pair of scalars~\cite{Cheung:2017yef,Cheung:2017ems}. 

We often work with particles $1$ and $n$ being scalars, so it is convenient to introduce a bar notation on the half-ladder numerators to indicate that we are considering a bi-scalar YM sector~\cite{Chen:2019ywi},
\begin{equation}\label{scalar_n_def}
\overline{N}^{(k)}(1,\ldots, n) \equiv \frac{\partial}{\partial \varepsilon_{1}{\cdot} \varepsilon_{n}}
N^{(k)} (1,\ldots, n) \, .
\end{equation}
The complete bi-scalar sector numerator is then ${\overline N}(1,\ldots, n) = \sum_{k=1}^{\lfloor n/2 \rfloor} {\overline N}^{(k)}(1,\ldots, n)$.

It is possible to invert the operation in~\eqn{scalar_n_def} and construct the YM numerators from the bi-scalar numerators~\cite{Chiodaroli:2017ngp,Chen:2019ywi},
\begin{equation}
\label{ym_n_from_scalar}
N^{(k)} (1,\ldots, n) = \frac{1}{k} \sum_{1\le i< j \le n} \varepsilon_{i}{\cdot} \varepsilon_{j} \overline{N}^{(k)} (i,\alpha_i, i+1, \ldots, j-1, \beta_j, j)\,,
\end{equation}
where $\alpha_i= [\cdots[[1,2],3],\ldots,i-1]$ and  $\beta_j= [j+1,[\ldots,n-2,[n-1,n]]\cdots]$ are nested commutators. 
Numerators of commutators distribute over their arguments to be consistent with the Lie-algebraic interpretation, $\overline{N}^{(k)} (\ldots,[X,Y],\ldots) \equiv \overline{N}^{(k)} (\ldots,X,Y,\ldots)-\overline{N}^{(k)} (\ldots,Y,X,\ldots)$. The boundary cases of the sum are handled through the identifications $\alpha_2=[1]=1$, $\beta_{n-1}=[n]=n$, and when either bracket is empty $\alpha_1=\beta_{n}=[\,]\rightarrow (-1)$ the numerator is multiplied by a minus sign. 

We can demonstrate \eqn{ym_n_from_scalar} using the four-point numerator as an example. The bi-scalar numerator is unique (given that $\overline{N}(1,2,3,4)=\overline{N}(4,3,2,1)$),
\begin{eqnarray} \label{biscalar4pt}
\overline{N}(1,2,3,4) = 
4 \varepsilon_2{\cdot} p_1 \varepsilon_3 {\cdot} p_1 + 4 \varepsilon_2{\cdot} p_1 \varepsilon_3 {\cdot} p_2 - 
 \varepsilon_2 {\cdot} \varepsilon_3 s_{12} \, ,
\end{eqnarray}
and can be further decomposed into polarization powers
\begin{align}
\overline{N}^{(1)}(1,2,3,4) &= 
4 \varepsilon_2{\cdot} p_1 \varepsilon_3 {\cdot} p_1 + 4 \varepsilon_2{\cdot} p_1 \varepsilon_3 {\cdot} p_2  \,, \nonumber \\
 \overline{N}^{(2)}(1,2,3,4)& = - 
 \varepsilon_2 {\cdot} \varepsilon_3 s_{12} \, .
\end{align}
Note that the polarization-power labels also keep track of the overall $\varepsilon_1 {\cdot} \varepsilon_4$ that was removed by the derivative $\frac{\partial}{\partial\varepsilon_1{\cdot}\varepsilon_n}$.
Using \eqn{ym_n_from_scalar} one can verify that 
\begin{align}
    N^{(k)}(1,2,3,4)
    =& \frac{1}{k}\Big[(\varepsilon_1{\cdot}\varepsilon_4) {\overline N}^{(k)}(1,2,3,4) -(\varepsilon_1{\cdot}\varepsilon_3) {\overline N}^{(k)}(1,2,4,3)-(\varepsilon_1{\cdot}\varepsilon_2) {\overline N}^{(k)}(1,[3,4],2) \nonumber  \\
 &\!\!-(\varepsilon_2{\cdot}\varepsilon_4) {\overline N}^{(k)}(2,1,3,4)
    +(\varepsilon_2{\cdot}\varepsilon_3) {\overline N}^{(k)}(2,1,4,3)-(\varepsilon_3{\cdot}\varepsilon_4) {\overline N}^{(k)}(3,[1,2],4)\Big]\nonumber \\ 
\end{align}
matches the numerator in \eqn{four_point_numerator}, given that $N_{[[1,2],3]}= N^{(1)}(1,2,3,4)+N^{(2)}(1,2,3,4)$.

For later purposes, it is convenient to introduce shorthand notation for the kinematic variables that frequently appear in the numerators. We define the variables\footnote{The $x_i$ are sometimes called ``region momenta'', and the $u_i$ are equivalent to cubic scalar-gluon vertices.}
\begin{equation}
\z_i^\mu \equiv \sum_{j=1}^i p_j^\mu \, ,\hspace{1cm} u_i \equiv 2\varepsilon_i\cdot \z_i \, ,
\end{equation}
which make the bi-scalar half-ladder numerators simpler to work with. For example, the four-point bi-scalar numerator (\ref{biscalar4pt}) now takes the simple form 
\begin{eqnarray}\label{4pt_n_uz}
\overline{N}(1,2,3,4) =u_2 u_3  -\varepsilon_2\cdot\varepsilon_3 \z_2^2 \,,
\end{eqnarray}
and the polarization-power components are then $\overline{N}^{(1)} = u_2 u_3$ and $\overline{N}^{(2)} =- \varepsilon_2\cdot\varepsilon_3 \z_2^2$. Note that this notation obscures the relabeling symmetries of the numerators so it should be used with some care, and we will mainly use it for bi-scalar half-ladder numerators.   

It is illuminating to attempt to represent the contributions to the numerators using diagrammatic notation. As a general guiding principle, we track inner products between polarization vectors using solid lines and wavy lines represent other contributions (typically inner products between polarizations and momenta). For example, the four-point bi-scalar numerators are represented by the two diagrams 
\begin{equation} \label{eq:4ptdiagrams}
    \overline{N}^{(1)} = \!\!
\begin{tikzpicture}[baseline={(0, 0.3cm)}]
\draw[thick] (-1.6,0) -- (1.6,0);
\draw[thick, snake it] (0.7,0) -- (0.7,1);
\draw[thick, snake it] (-0.7,0) -- (-0.7,1);
\node at (-1.8,0) {$1$};
\node at (-0.7,1.2) {$2$};
\node at (0.7,1.2) {$3$};
\node at (1.8,0) {$4$};
\end{tikzpicture} \!= u_2 u_3
\, ,
\hspace{0.5cm}
    \overline{N}^{(2)} = \!\!
\begin{tikzpicture}[baseline={(0, 0.3cm)}]
\draw[thick] (-1.6,0) -- (1.6,0);
\draw[thick] (-0.7,0.05) -- (0.7,0.05);
\draw[thick] (0.7,0.05) -- (0.7,1);
\draw[thick] (-0.7,0.05) -- (-0.7,1);
\node at (-1.8,0) {$1$};
\node at (-0.7,1.2) {$2$};
\node at (0.7,1.2) {$3$};
\node at (1.8,0) {$4$};
\end{tikzpicture} \!= 
- \varepsilon_2\cdot\varepsilon_3 \, \z_2^2
\, .
\end{equation}
This notation makes it clear that the numerator $\overline{N}^{(2)}$ represents a contribution coming from an auxiliary rank-2 field in the intermediate channel. Indeed, we will see in section~\ref{NMHVsection} that this interpretation is correct. 

At five points, the half-ladder BCJ numerator can again be decomposed into two bi-scalar numerators, $\overline{N}^{(1)}$ and $\overline{N}^{(2)}$, of different polarization powers. Focusing on $\overline{N}^{(2)}$, we realize even before having an explicit formula that there are three different dot products $\ee{3}{4}\, ,\ee{2}{3}$ and $\ee{2}{4}$, which can be represented by the schematic diagrams 
\begin{eqnarray} \label{eq:5ptdiagrams}
\overline{N}^{(2)}(1,2,3,4,5) &=& 
\begin{tikzpicture}[baseline={(0, +0.3cm)}]
\draw[thick] (-2,-0.05) -- (2,-0.05);
\draw[thick] (0,0) -- (1,0);
\draw[thick] (0,0) -- (0,1);
\draw[thick] (1,0) -- (1,1);
\draw[thick,snake it] (-1,-0.05) -- (-1,1);
\node at (-2.1,-0.05) {$1$};
\node at (-1,1.2) {$2$};
\node at (0,1.2) {$3$};
\node at (1,1.2) {$4$};
\node at (2.1,-0.05) {$5$};
\end{tikzpicture} 
+
\begin{tikzpicture}[baseline={(0, +0.3cm)}]
\draw[thick] (-2,-0.05) -- (2,-0.05);
\draw[thick] (-1,0) -- (0,0);
\draw[thick] (0,0) -- (0,1);
\draw[thick,snake it] (1,-0.05) -- (1,1);
\draw[thick] (-1,0) -- (-1,1);
\node at (-2.1,-0.05) {$1$};
\node at (-1,1.2) {$2$};
\node at (0,1.2) {$3$};
\node at (1,1.2) {$4$};
\node at (2.1,-0.05) {$5$};
\end{tikzpicture} \notag
\\
&& +
\begin{tikzpicture}[baseline={(0, +0.3cm)}]
\draw[thick] (-2,-0.05) -- (2,-0.05);
\draw[thick] (-1,0) -- (1,0);
\draw[thick,snake it] (0,0) -- (0,1);
\draw[thick] (1,0) -- (1,1);
\draw[thick] (-1,0) -- (-1,1);
\node at (-2.1,-0.05) {$1$};
\node at (-1,1.2) {$2$};
\node at (0,1.2) {$3$};
\node at (1,1.2) {$4$};
\node at (2.1,-0.05) {$5$};
\end{tikzpicture} \,.
\end{eqnarray}
The explicit expressions for these diagrams will be given in section~\ref{NMHVsection}.

Likewise at $n$ points, the half-ladder bi-scalar numerator $\overline{N}^{(2)}$ can be schematically written down as a sum over $(n-2)(n-3)/2$ different diagrams, 
\begin{equation}\label{eq_pp2_numerator_schematic}
\overline{N}^{(2)}(1,\ldots,n) = 
\sum_{1<i<j<n}
\begin{tikzpicture}[baseline={(0, +0.3cm)}]
\draw[thick] (-3,0) -- (3,0);
\draw[thick] (-0.95,0.05) -- (0.95,0.05);
\draw[thick,snake it] (-2.2,0) -- (-2.2,1);
\draw[thick,snake it] (-0.5,0.05) -- (-0.5,1);
\draw[thick,snake it] (0.5,0.05) -- (0.5,1);
\node at (-1.8,0.5) {$\ldots$};
\node at (0,0.5) {$\ldots$};
\draw[thick,snake it] (-1.4,0) -- (-1.4,1);
\draw[thick] (0.95,0.05) -- (0.95,1);
\draw[thick] (-0.95,0.05) -- (-0.95,1);
\draw[thick,snake it] (2.2,0) -- (2.2,1);
\node at (1.8,0.5) {$\ldots$};
\draw[thick,snake it] (1.4,0) -- (1.4,1);
\node at (-3.1,0) {$1$};
\node at (3.2,0) {$n$};
\node at (0.95,1.2) {$j$};
\node at (-0.95,1.2) {$i$};
\end{tikzpicture} 
\, ,
\end{equation}
and in section~\ref{NMHVsection} we will find an explicit formula for the diagrams using a Lagrangian. Before considering these NMHV amplitude contributions, we will have a short section on the Lagrangian that generates the ${N}^{(1)}$ numerators which gives MHV amplitudes.

\section{MHV Lagrangian\label{MHVsection}}
Tree-level MHV amplitudes of YM can be computed from the BCJ numerators with the lowest polarization power $N^{(1)}$~\cite{Chen:2019ywi}. For example, if legs $1$ and $2$ carry negative helicity and the remaining legs have positive helicity, then choosing reference null momenta $q_i^\mu$ as
\begin{equation} \label{pol_cond}
    \varepsilon_1^{-}(p_1, q_1=p_2)\,,~~~~~ \varepsilon_2^{-}(p_2, q_2=p_1)\,,~~~~~\varepsilon_{i>2}^{+}(p_i, q_i=p_1)\,
\end{equation}
ensures that the only non-vanishing polarization products are $\varepsilon_2 \cdot \varepsilon_i$. Because they all contain $\varepsilon_2$ these factors can at most appear linearly in the numerator, hence $N^{(1)}$ is sufficient. 

By dimensional analysis the numerators $N^{(1)}$ cannot contain products of momenta $p_i \cdot p_j$ hence they cannot give rise to inverse propagators $p^2 \sim \Box$, and thus there are no hidden contact terms inside the numerators. This implies that $N^{(1)}$ must originate from a YM Lagrangian that neither has contact terms nor auxiliary fields. Thus we can simply truncate the YM Lagrangian\footnote{We assume Lorenz gauge $\partial \cdot A=0$ (or, more accurately, Feynman gauge) since the BCJ numerators are defined to sit on top of propagators that are in Feynman gauge, see \eqn{prop_denominator}.} to cubic order, which defines the MHV Lagrangian 
\begin{equation} \label{MHVLagr}
    \mathcal{L}_{\textrm{MHV}} = \mathcal{L}_{\textrm{YM}}\Big|_{\rm cubic} = \text{Tr} \, \left( \frac{1}{2} A_\mu \Box A^\mu -\partial_\mu A_\nu [A^\mu,A^\nu] \right) \, .
\end{equation}
For simplicity, we have set the gauge coupling to unity, $g=1$, which we will do henceforth.

The Lagrangian \eqref{MHVLagr} gives standard cubic YM Feynman rules that produce numerators of all polarization power sectors $k\ge 1$, but for the purpose of this section we can assume that the resulting numerators are truncated to $N^{(1)}$ because of the conditions \eqref{pol_cond} imposed on the polarization vectors. That this Lagrangian is invalid beyond this sector is obvious from the fact that the truncation of the quartic term is not a valid (gauge-fixing) operation. 

Let us now show that the numerators $N^{(1)}$ computed from the Lagrangian \eqref{MHVLagr} obey color-kinematics duality. We use the diagrammatic notation from the previous section to make the argument simpler. The $N^{(1)}$ numerators are computed from sums of diagrams each having just one solid line (representing the contracted polarization vectors). We can think of the solid line as a scalar line, and we have to sum over all possible pairs of external states that are joined by this line. The scalar line can be identified from the equations of motion of the MHV Lagrangian,
\begin{eqnarray}\label{cubic_ym_truncated}
\Box A^\mu &= -2[A_\nu, \partial^\nu A^\mu] + [A_\nu , \partial^\mu A^\nu] \, ,
\end{eqnarray}
where the second term produces contractions that ultimately connect external polarization vectors. The first term corresponds to cubic interactions were no scalar line is present, thus we think of it as consisting of only wavy lines. It is straightforward to then study the triplet sum of off-shell numerators obtained using these two interactions. One can show that the equation
\begin{equation} \label{eq:4ptdiagrams_jacobi}
\begin{tikzpicture}[baseline={(0, 0.3cm)}]
\draw[thick] (-1.6,0) -- (1.6,0);
\draw[thick, snake it] (0.7,0) -- (0.7,1);
\draw[thick, snake it] (-0.7,0) -- (-0.7,1);
\node at (-1.8,0) {$1$};
\node at (-0.7,1.2) {$2$};
\node at (0.7,1.2) {$3$};
\node at (1.8,0) {$4$};
\end{tikzpicture} 
-
\begin{tikzpicture}[baseline={(0, 0.3cm)}]
\draw[thick] (-1.6,0) -- (1.6,0);
\draw[thick, snake it] (0.7,0) -- (0.7,1);
\draw[thick, snake it] (-0.7,0) -- (-0.7,1);
\node at (-1.8,0) {$1$};
\node at (-0.7,1.2) {$3$};
\node at (0.7,1.2) {$2$};
\node at (1.8,0) {$4$};
\end{tikzpicture} 
=
\begin{tikzpicture}[baseline={(0, 0.3cm)}]
\draw[thick] (-0.7,0) -- (0.7,0);
\draw[thick, snake it] (0.7,1.1) -- (0.0,0.7);
\draw[thick, snake it] (0.0,0.7) -- (-0.7,1.1);
\draw[thick, snake it] (0,0) -- (0.0,0.7);
\node at (-0.9,0) {$1$};
\node at (-0.8,1.3) {$2$};
\node at (0.8,1.3) {$3$};
\node at (0.9,0) {$4$};
\end{tikzpicture} 
\, ,
\end{equation}
holds up to terms proportional to $\varepsilon_2{\cdot}p_2$ and $\varepsilon_3{\cdot}p_3$. Such terms vanish on-shell because of the transversality of $\varepsilon_i$, but we need to show that the above three-term identity holds for a generic off-shell numerator embedded into a larger tree diagram. Thus we need to show that transversality holds for all off-shell wavy lines in the MHV sector. 

We can show this using induction, essentially feeding an arbitrary number of wavy-line interactions into~\eqn{eq:4ptdiagrams_jacobi}. Suppose that to some wavy line $i$ we attach a cubic diagram generated from the first interaction term in~\eqn{cubic_ym_truncated}, producing two new wavy lines $j,k$, so $i\to[jk]$. Then its effective polarization vector\footnote{Note that this is not to be confused with an external properly normalized polarization vector.} takes the form
\begin{equation}\label{diffeo_current}
    \varepsilon_{i}^\mu = -2 ( \varepsilon_j{\cdot}p_k \varepsilon_k^\mu - \varepsilon_k{\cdot}p_j \varepsilon_j^\mu ) \, ,
\end{equation}
in terms of the effective polarizations of the lines $j,k$. 
 By the induction hypothesis, we assume that the latter polarizations are transverse, $\varepsilon_j{\cdot}p_j = \varepsilon_k{\cdot}p_k=0$. Using this, and the antisymmetry in the labels $j$ and $k$, \eqn{diffeo_current} implies that $\varepsilon_i{\cdot}p_i = 0$. Thus the wavy-line interaction preserves transversality, and since the wavy lines will eventually terminate in proper external states  at tree level the induction hypothesis is correct.  This completes the argument that \eqn{eq:4ptdiagrams_jacobi} holds off shell. 

We also need to consider kinematic Jacobi identities of diagrams containing only the first term in~\eqn{cubic_ym_truncated}, which is the wavy-line interaction~\eqref{diffeo_current}. However, note that \eqn{diffeo_current} is essentially a momentum-space Lie bracket of plane-wave\footnote{We suppress the plane-wave color factor as well as overall imaginary factors.} vector fields $A_j^\mu=\varepsilon_j^\mu e^{i  p_j \cdot x}$ ,
\begin{equation}
A_i \cdot \partial = -2 \, \big[A_j \cdot \partial, \, A_k \cdot \partial \, \big]\,,
\end{equation}
and so it automatically obeys the Jacobi identity. Since the vector fields are transverse $\partial \cdot A_i=0$, we have thus exposed a kinematic sub-algebra that corresponds to volume-preserving diffeomorphisms (in analogy with the previously found 2D area-preserving~\cite{Monteiro:2011pc} and 3D volume-preserving diffeomorphisms~\cite{Ben-Shahar:2021zww}). However, we cannot reproduce the full $N^{(1)}$ numerators from commutators of only these diffeomorphism generators, since the full numerator is a superposition of all diagrams with a single solid line going between pairs of external states. Nevertheless, by linearity of the kinematic Jacobi identity, the superposition of all such diagrams will give $N^{(1)}$ numerators that obey color-kinematics duality.  

To summarize, for tree level diagrams obtained from the standard cubic YM Lagrangian, color-kinematics duality is satisfied at polarization power one, or equivalently, for MHV amplitudes. Note that the off-shell argument does not automatically extend to loop level since it was important that the recursive transversality argument terminates with external on-shell legs, which is always true at tree level, but not at loop level. 

Finally, let us give the half-ladder bi-scalar numerators that are generated from the Feynman rules of the MHV Lagrangian~\eqref{MHVLagr}. They are simply 
\begin{equation}\label{eq_pp1_numerator_schematic}
\overline{N}^{(1)}(1,2\ldots,n-1,n) = 
\begin{tikzpicture}[baseline={(0, +0.3cm)}]
\draw[thick] (-1.4,0) -- (3,0);
\draw[thick,snake it] (-0.5,0) -- (-0.5,1);
\draw[thick,snake it] (0.4,0) -- (0.4,1);
\node at (0.9,0.5) {$\ldots$};
\draw[thick,snake it] (2.2,0) -- (2.2,1);
\draw[thick,snake it] (1.4,0) -- (1.4,1);
\node at (-1.5,0) {$1$};
\node at (3.2,0) {$n$};
\node at (-0.5,1.2) {$2$};
\node at (2.2,1.2) {$n-1$};
\end{tikzpicture} = \, \prod_{i=2}^{n-1} u_i
\, ,
\end{equation}
and the remaining contributions to the half-ladder numerator ${N}^{(1)}$ can be obtained using \eqn{ym_n_from_scalar}, or from the Feynman rules of the Lagrangian \eqref{MHVLagr}. The non-half-ladder diagrams are similarly computed either from commutators of the  half-ladder numerator ${N}^{(1)}$, or from the same Feynman rules. 

\section{NMHV Lagrangian\label{NMHVsection}}
In this section we construct several cubic Lagrangians that produce BCJ numerators at polarization power two, which allow us to compute helicity amplitudes up to the NMHV sector of YM. We begin the construction at four points, and work our way upwards in multiplicity by adding terms to the Lagrangian, until the corrections terminate at seven points. 

\subsection{The four-point NMHV Lagrangian}
Let us start from the standard YM Lagrangian (subject to Lorenz gauge $\partial \cdot A=0$),
\begin{equation} \label{eq:ymlag}
\mathcal{L}_{\text{YM}}=- \frac{1}{4} \text{Tr} \,  (F^{\mu \nu})^2 = \text{Tr} \, \left( \frac{1}{2} A_\mu \Box A^\mu -\partial_\mu A_\nu [A^\mu,A^\nu] - \frac{1}{4} [A_\mu,A_\nu][A^\mu,A^\nu] \right) \,  ,
\end{equation}
where the field strength is $F^{\mu \nu}= \partial^\mu A^\nu -\partial^\nu A^\mu + [A^\mu , A^\nu] $.
For later purposes, let us also quote the corresponding equations of motion 
\begin{equation}
    \Box A^\nu = -2 [A_\mu, \partial^\mu A^\nu]+[A_\mu, \partial^\nu A^\mu] - [A_\mu, [A^\mu, A^\nu]]\, .
\end{equation}
 We seek to find a cubic rewriting of this Lagrangian that generates BCJ numerators at polarization power two. As a first step, we can resolve the quartic interaction by introducing propagating auxiliary fields. There are several ways to do this.  We find that an elegant solution is to reinterpret the nested commutator, appearing in the last term of the above equations of motion, to contain a two-form tensor $B^{\mu \nu} = -1/2 [A^\mu, A^\nu]$. In order to allow for a suitable kinetic term of correct mass dimension, we introduce a companion field $\tilde{B}^{\mu \nu}$ of mass dimension zero, into which $B^{\mu \nu}$ propagates. The Lagrangian we obtain is
\begin{equation}
\label{eq:bfieldym}
    \mathcal{L}_4 = \text{Tr} \, \left( \frac{1}{2} A_\mu \Box A^\mu -\partial_\mu A_\nu [A^\mu,A^\nu] + B_{\mu \nu} \Box \tilde{B}^{\mu \nu} + \frac{1}{2}  [A_\mu , A_\nu] (B^{\mu \nu} + \Box \tilde{B}^{\mu \nu}) \right),
\end{equation}
and the equations of motion for $B^{\mu \nu}$ and $\tilde{B}^{\mu \nu}$ are given by
\begin{equation}
    B^{\mu \nu} = \Box \tilde{B}^{\mu \nu} =-\frac{1}{2} [A^\mu,A^\nu]\, ,
\end{equation}
which we can plug into the Lagrangian above to recover $\mathcal{L}_{\text{YM}}$ as in \eqn{eq:ymlag}.
To compute amplitudes from this Lagrangian one has to restrict the external legs to be vectors $A^\mu$, since the $B$ and $\tilde{B}$ fields are composite (the linearized excitations of $B$ vanish). As discussed in the previous section, the cubic-in-$A^\mu$ interactions of this Lagrangian correctly reproduce the MHV sector of YM, while the double line in the polarization-power two sector (see for example the second graph in \eqn{eq:4ptdiagrams}) corresponds to propagation of the $B^{\mu \nu}$ and $\tilde{B}^{\mu \nu}$ fields. Since these fields can be integrated out yielding the standard YM Lagrangian, the Feynman rules of \eqn{eq:bfieldym} reproduce standard YM tree amplitudes. 
At four points, the BCJ numerator is unique~\cite{Chen:2021chy} (up to an overall normalization) upon imposing relabeling symmetry and reflection symmetry, as given in \eqn{four_point_numerator}, and it is also correctly reproduced by the Lagrangian \eqref{eq:bfieldym}. In fact, any cubic action for YM theory would reproduce the correct four-point numerator, but we find that the above choice of ${\cal L}_4$ with both the $B$ and $\tilde{B}$ fields allows for the necessary freedom to proceed to generating BCJ numerators at higher points. 

\subsection{The five-point NMHV Lagrangian}
Going to five points, we can immediately see that $\mathcal{L}_4$ is unable to generate half-ladder numerators $N^{(2)}$  proportional to $\varepsilon_1 {\cdot} \varepsilon_5 \, \varepsilon_2 {\cdot} \varepsilon_4$, which is the last diagram pictured in \eqn{eq:5ptdiagrams}, and corresponds to the double-line field emitting a gluon line. To generate this kind of graph we need new interaction terms of the schematic form $\partial A B \tilde{B}$. In principle, introducing new interactions could change the equations of motion, change the four-point amplitude, and break gauge-invariance of YM theory. Thus, one can expect it to be a delicate procedure.  

Let consider adding new linear-in-$B^{\mu \nu}$ or linear-in-$\tilde{B}^{\mu \nu}$ terms on the right-hand side of their corresponding equations of motion. By repeated insertion of the equations of motion into themselves, we would generate a set of non-local higher order $A^\mu$ interactions (similar to refs.~\cite{Bern:2010yg,Tolotti:2013caa}). Can we make sure that such new interactions always cancel out in any tree-level amplitude computation? Let us start, for the sake of simplicity, by requiring something more specific, namely that the deformations introduced by the Lagrangian terms $\partial AB \tilde{B}$ vanish within the subsystem of equations of motion for the auxiliary fields. This can be easily achieved if they take the following form
\begin{equation} \label{beom1}
    B^{\mu \nu} = -\frac{1}{2} [A^\mu,A^\nu] + \alpha \frac{\partial_\rho}{\Box} \left( [A^\mu , B^{\rho \nu}] + \text{cyclic}(\mu \rho \nu) \right),
\end{equation}
where $\alpha$ is a free parameter for now. When these equations of motion are repeatedly substituted back into themselves $n$ times, the sum over cyclic permutations of Lorentz indices in the commutator manifestly vanishes by Jacobi identity at any order in $A$, yielding
\begin{equation} \label{beom2}
    B^{\mu\nu,(n)} = -\frac{1}{2}[A^\mu,A^\nu] + {\cal O}(\alpha^n) \, .
\end{equation}
where the last term is ${\cal O}(\alpha^n) \sim \alpha^n (A^\mu)^{n} B^{\nu \rho,(0)}$, and it vanishes in perturbative tree-level computations since we take $B^{\nu \rho,(0)}=0$ for external states. 
The equations of motion in \eqn{beom1} must arise from introducing the following interactions in the Lagrangian:
\begin{equation}
\Delta \mathcal{L}_5 = - \alpha \, \text{Tr}  \,  \partial_\rho \tilde{B}_{\mu \nu} \Big( [A^\mu, B^{\nu \rho}] + \text{cyclic}(\mu \nu \rho) \Big) .
\end{equation}
where $ \mathcal{L}_5=  \mathcal{L}_4 + \Delta  \mathcal{L}_5 $ is the total Lagrangian. As argued, the term in parentheses does not contribute to on-shell tree-level scattering amplitudes by the Jacobi identity, but it does alter the individual $N^{(2)}$ numerators. 

With the new Lagrangian $ \mathcal{L}_5$ the equations of motion for $\tilde{B}^{\mu \nu}$ now read
\begin{equation}
 \tilde{B}^{\mu \nu} =   - \frac{1}{2}\frac{1}{\Box} [A^\mu, A^\nu] + \alpha\frac{1}{\Box} \big[A_\rho\,,\, \partial^\mu \tilde{B}^{\rho \nu} + \text{cyclic}(\mu \rho \nu)\big] .
 \end{equation}
Note that repeated insertion of the $\tilde{B}$-field equations of motion into themselves generates an infinite set of contact terms of increasing $A$ order.
However, we can ignore this, since from using the $B$-equation alone it follows that the standard YM Lagrangian is recovered (up to terms that do not contribute to tree-level amplitudes).

We now need to probe whether the deformed numerators enjoy color-kinematics duality for some choice of $\alpha$. We can do so by explicitly checking Jacobi relations between the $N^{(2)}$ numerators, or checking that the DDM-decomposed partial amplitude~\eqref{eq:BAampli} returns a gauge invariant quantity. At five points, we confirm that  BCJ numerators are obtained after fixing the free parameter in $\Delta \mathcal{L}_5$ to the value $\alpha=2$, giving the duality-satisfying Lagrangian 
\begin{align}
\label{eq:l5}
    \mathcal{L}_5 &= \text{Tr} \, \Big( \frac{1}{2} A_\mu \Box A^\mu -\partial_\mu A_\nu [A^\mu,A^\nu] + B_{\mu \nu} \Box \tilde{B}^{\mu \nu} + \frac{1}{2}  [A_\mu , A_\nu] (B^{\mu \nu} + \Box \tilde{B}^{\mu \nu}) \notag  \\ 
    & \quad \quad \quad  +   4 \partial_\nu \tilde{B}_{\mu \rho} [A^\mu ,B^{\nu \rho}] - 2 \partial_\rho \tilde{B}_{\mu \nu} [ A^\rho,  B^{\mu \nu}] \Big)\,.
\end{align}

We thus find the following NMHV contributions to the bi-scalar sector diagrams that we introduced in section~\ref{sec:preliminaries},
\begin{eqnarray} 
\begin{tikzpicture}[baseline={(0, +0.3cm)}]
\draw[thick] (-2,-0.05) -- (2,-0.05);
\draw[thick] (0,0) -- (1,0);
\draw[thick] (0,0) -- (0,1);
\draw[thick] (1,0) -- (1,1);
\draw[thick,snake it] (-1,-0.05) -- (-1,1);
\node at (-2.1,-0.05) {$1$};
\node at (-1,1.2) {$2$};
\node at (0,1.2) {$3$};
\node at (1,1.2) {$4$};
\node at (2.1,-0.05) {$5$};
\end{tikzpicture} &~~=~~&  \varepsilon_3 \cdot \varepsilon_4 (\varepsilon_2{\cdot} \z_3 \, \z_2^2 - u_2 \, \z_3^2 ) \, ,
\\
\begin{tikzpicture}[baseline={(0, +0.3cm)}]
\draw[thick] (-2,-0.05) -- (2,-0.05);
\draw[thick] (-1,0) -- (1,0);
\draw[thick,snake it] (0,0) -- (0,1);
\draw[thick] (1,0) -- (1,1);
\draw[thick] (-1,0) -- (-1,1);
\node at (-2.1,-0.05) {$1$};
\node at (-1,1.2) {$2$};
\node at (0,1.2) {$3$};
\node at (1,1.2) {$4$};
\node at (2.1,-0.05) {$5$};
\end{tikzpicture} &~~=~~&  -\frac{1}{2}\varepsilon_2 \cdot \varepsilon_4 u_3 ( \z_2^2 + \z_3^2 ) \, .
\end{eqnarray}
As before, the remaining $N^{(2)}$ contributions to the half-ladder diagrams can be obtained using \eqn{ym_n_from_scalar}, or from the Feynman rules of the Lagrangian~\eqref{eq:l5}. Since we are at five points, all cubic diagrams are of the half-ladder type, and thus this completes the five-point construction. 

Before proceeding to discuss the six-point case, let us point out a remarkable fact about the Lagrangian~\eqref{eq:l5}. We have checked through multiplicity $n=11$ that its Feynman rules correctly compute half-ladder BCJ numerators in the bi-scalar sector $\overline N^{(2)}$. Inserting those into~\eqn{ym_n_from_scalar}, gives all half-ladder numerators $N^{(2)}$, and further use of Jacobi relations gives all non-half-ladder numerators in the NMHV sector, up to the multiplicity we checked. Based on this rigorous pattern, we conjecture that the Lagrangian~\eqref{eq:l5} correctly computes the bi-scalar NMHV numerators $N^{(2)}$ to any multiplicity at tree level, and furthermore implicitly provides a good BCJ representation for the remaining diagrams in this sector. 

\subsection{Closed form representation of the bi-scalar sector numerators}
Let us summarize what we have achieved thus far by giving a closed form expression for the bi-scalar numerators. As shown in \eqn{eq_pp1_numerator_schematic}, the MHV bi-scalar numerators take a simple form
\begin{eqnarray}
\overline{N}^{(1)}(1,2,\ldots,n-1,n) =  u_2 u_3\ldots u_{n-1} \, .
\end{eqnarray}
Next, we gave a schematic diagram representation for the bi-scalar sector NMHV numerators in~\eqn{eq_pp2_numerator_schematic}, and with the Lagrangian $\mathcal{L}_5$ in \eqn{eq:l5} we can make this more precise. Since we are considering numerators at polarization power two, each contribution must have exactly one vertex of type $AAB$ and one of type $AA\square\tilde{B}$ in the diagram. Consider a half-ladder diagram where the $B$ field is first sourced at position $i$ to the left of the final $\tilde{B}$ field at position $k$. Between these two vertices a chain of $AB\partial\tilde{B}$-type vertices is inserted, giving the below diagram 
\begin{equation}
\mathcal{D}_{ijkn} = 
\begin{tikzpicture}[baseline={(0, +0.3cm)}]
\draw[thick] (-3,0) -- (3.5,0);
\draw[thick] (-0.95,0.05) -- (0.95,0.05);
\draw[thick,snake it] (-2.2,0) -- (-2.2,1);
\draw[thick,snake it] (-0.5,0.05) -- (-0.5,1);
\draw[thick,snake it] (0.5,0.05) -- (0.5,1);
\node at (-1.8,0.5) {$\ldots$};
\node at (0,0.5) {$\ldots$};
\draw[thick,snake it] (-1.4,0) -- (-1.4,1);
\draw[thick] (0.95,0.05) -- (0.95,1);
\draw[thick] (-0.95,0.05) -- (-0.95,1);
\draw[thick,snake it] (2.2,0) -- (2.2,1);
\node at (1.8,0.5) {$\ldots$};
\draw[thick,snake it] (1.4,0) -- (1.4,1);
\draw[thick,snake it] (3,0) -- (3,1);
\node at (2.6,0.5) {$\ldots$};
\node at (-3.1,0) {$1$};
\node at (3.7,0) {$n$};
\node at (0.95,1.2) {$j$};
\node at (-0.95,1.2) {$i$};
\node at (2.2,1.2) {$k$};
\node at (2,-0.2) {$\Box$};
\end{tikzpicture} 
\, ,
\end{equation}
where $j$ labels a scalar line (which need not be at position $k$), and the $\Box$ indicates the location of the inverse propagator $\z_{k-1}^2$.
The diagram where $\tilde{B}$ is sourced first can be obtained from reflection of this diagram. 
Since a solid line stretches all the way between $i$ and $j$, the intermediate vertices must be of the form $A^\mu B_{\nu\rho}\partial_\mu B^{\nu\rho}$, which gives a product of $u_l$ variables. Vertices between legs $j$ and $k$ can either be of the same type, or of the type $A^\mu B^{\nu\rho}\partial_\nu\tilde{B}_{\mu\rho}$.
The vertices before $i$ or after $k$ are of the type $AA\partial A$ and can only give rise to $u_l$ variables in order to not increase the polarization power. Thus, all together, each diagram is given by the expression
\begin{equation}
\mathcal{D}_{ijkn} =  \ee{i}{j} \z_{k-1}^2 U_{2,i-1}U_{i+1,j-1} \big(\z_{j-1}\cdot V_{j+1}\cdots V_{k-1}\cdot \varepsilon_k\big) U_{k+1,n-1} \, ,
\end{equation}
where the matrices $(V_i)^{\mu\nu} $ are given by
\begin{eqnarray}
(V_i)^{\mu\nu} = u_i \eta^{\mu\nu}  - 2\varepsilon_{i}^\mu \z_{i-1}^\nu \, ,
\end{eqnarray}
and the $U$'s are products of consecutive $u_i$, defined by
\begin{equation} \label{Udef}
    U_{i,j} = u_i u_{i+1}\cdots u_{j} \, .
\end{equation}
We set the boundary case $j=k$ of the expression in parenthesis $(\z_{j-1}\cdot V_j\cdots V_{k-1}\cdot \epsilon_k)$ to be equal to the number $(-1/2)$ since otherwise the polarization vector $\varepsilon_{k}$ is double counted. Finally, using the diagrams $\mathcal{D}_{ijkn}$ the bi-scalar numerator in this  sector is
\begin{eqnarray} \label{N2biScalarnumerator}
\overline{N}^{(2)}(1,\ldots,n) = \sum_{1<i<j\leq k}^{n-1} \mathcal{D}_{ijkn} + \textrm{reflection} \, .
\end{eqnarray}
The reflected diagram is obtained by reversing the labels $\{1,2,\dots,n\}$ on the momenta $p_i$ and polarizations $\varepsilon_i$, not the region momenta $x_i$.
It is interesting to note that the matrices $V_i$ play a similar role to the $G^i{}_j$ matrices used for NMHV numerators in ref.~\cite{Chen:2021chy}, except that the former depend on region momenta $x_i$ and the latter on particle momenta $p_i$. 

\subsection{NMHV Lagrangian beyond the bi-scalar sector}
As detailed in the previous two subsections, we conjectured that the bi-scalar sector numerators~\eqref{N2biScalarnumerator} generated by the five-point Lagrangian $\mathcal{L}_5$ \eqref{eq:l5} give valid BCJ numerators to all multiplicity at tree level. From these we can then uniquely compute all NMHV contributions beyond the bi-scalar sector.  We now wish to find a Lagrangian description of such contributions, which starts at six points.   

First, let us write down an explicit expression for the six-point bi-scalar numerator (summing over polarization power one and two),
\begin{align}
\overline{N}(1,2,3,4,5,6)&= U_{2,5} +   ( \mathcal{D}_{2336} + \mathcal{D}_{2346}+  \mathcal{D}_{2356}+  \mathcal{D}_{2446}+
 \mathcal{D}_{2456}+  \mathcal{D}_{2556}+ \mathcal{D}_{3446} \notag \\
 & \quad +  \mathcal{D}_{3456}+  \mathcal{D}_{3556}+  \mathcal{D}_{4556}  + \text{reflections}) \notag \\
\quad &= u_2 u_3 u_4 u_5 -2 \ee{2}{3} \left( 2 \ez{4}{2} \ez{5}{3} \z_4^2 - \ez{5}{2}   u_4 \z_4^2 - \ez{4}{2}  u_5 \z_3^2 +  u_4 u_5 \z_2^2 \right) \notag \\
& \quad + \ee{2}{4} \left( 2 \ez{5}{3}  u_3 \z_4^2 - (\z_2^2 + \z_3^2) u_3 u_5 \right)  \notag \\
& \quad - \ee{2}{5} \left( \z_2^2+ \z_4^2 \right) u_3 u_4 \notag \\
& \quad +2 \ee{3}{4} \left( \ez{5}{3} u_2 \z_4^2 + \ez{2}{3}  u_5 \z_2^2 -  u_2 u_5 \z_3^2 \right) \notag \\
& \quad + \ee{3}{5} \left( 2 \ez{2}{3}  u_4 \z_2^2 -(\z_4^2+\z_3^2) u_2 u_4 \right) \notag \\
& \quad -2 \ee{4}{5} \left(2 \ez{2}{3} \ez{3}{4} \z_2^2 - \ez{3}{4}  u_2 \z_3^2 - \ez{2}{4}  u_3 \z_2^2 +  u_2 u_3 \z_4^2\right).
\end{align}
Plugging this numerator into the formula \eqref{ym_n_from_scalar} gives the remaining NMHV numerator contributions $N^{(2)}$ at six points, which do not match (diagram-by-diagram) the same contributions computed from the Lagrangian $\mathcal{L}_5$. 
It is clear that the numerators obtained from~\eqn{ym_n_from_scalar} are the ones we need since they satisfy color-kinematics duality, thus we must modify the Lagrangian $\mathcal{L}_5$ with new six-point contributions.  
 
 We proceed by looking at individual kinematic monomials in $N^{(2)}$ that pinpoint the mismatch, and then infer what are the simplest interactions that can restore color-kinematics duality. For example, considering the term $u_3 \ee{2}{4} \ee{5}{6}  \varepsilon_3 {\cdot} (p_5 - p_6) \z_2^2$,  we find that it is necessary to introduce a pair of new vector fields $Z^\mu$ and $\tilde{Z}^\mu$ of mass dimension one, which interact with $A^\mu$, $B^{\mu \nu}$ and $\tilde{B}^{\mu \nu}$. To ensure that the new auxiliary vectors do not pollute the four and five-point construction, we must constrain certain interactions. Specifically, we will only allow an interaction of the form $AAZ$ but no conjugate $AA\tilde{Z}$ interaction, thus ensuring that at most the $Z$ field can be sourced at four points, and hence it cannot propagate to a $\tilde{Z}$. Furthermore, we require that the $Z$, $\tilde{Z}$ fields can only source the linear combination of the tensor field $( B^{\mu \nu} -  \Box \tilde{B}^{\mu \nu} )$, which is specifically not sourced by $A^\mu$ fields at five points, hence the $Z$, $\tilde{Z}$ field cannot propagate at five points.

Let us see how this example plays out in detail. Adding two interactions of the form $  A^\mu\tilde{Z}^\nu B_{\mu\nu}$ and $  \partial_\nu A^\mu A_\mu  Z^\nu $ generates a new diagram contribution of the schematic form
\begin{equation}
    u_3 \, \ee{2}{4} \, \ee{5}{6}\, \varepsilon_1{\cdot}(p_5-p_6)  x_2^2   \ \ \ \longrightarrow \ \ \  
\label{eq:diag1}
\begin{tikzpicture}[baseline={(0, +0.3cm)}]
\draw[thick] (-2.5,0) -- (1.15,0);
\draw[thick] (-1.5,1) -- (-1.5,0.05) -- (0.5,0.05) -- (0.5,1);
\draw[thick] (1.5,1) -- (1.5,0) -- (2.5,0);
\draw[thick, snake it] (-0.5,0.05) -- (-0.5,1);
\node at (-2.6,0) {$1$};
\node at (-1.5,1.2) {$2$};
\node at (-.5,1.2) {$3$};
\node at (.5,1.2) {$4$};
\node at (1.5,1.2) {$5$};
\node at (2.6,0) {$6$};
\node at (1.32,0.05) {$\partial$};
\node at (-1.4,-0.3) {$\square \tilde{B}$};
\node at (-0.5,-0.3) {$B\tilde{B}$};
\node at (0.5,-0.3) {$B\tilde{Z}$};
\node at (1.5,-0.35) {$Z$};
\end{tikzpicture} \, , 
\end{equation}
where we have placed the previously mentioned offending monomial to the left and the corresponding schematic diagram with new vertices $A \tilde{Z}B$ and $AAZ$ to the right. The solid line that ends on a derivative indicates the  $\varepsilon_1{\cdot}(p_5-p_6)$  contraction.

Similarly,  flipping the tilde and non-tilde fields yields a correspondence between the following offending monomial and new diagram
\begin{align}
    u_3 \, \ee{1}{4} \, \ee{5}{6} \,  \varepsilon_2{\cdot}(p_5-p_6) \z_3^2 
\ \ \ \longrightarrow \ \ \  
\begin{tikzpicture}[baseline={(0, +0.3cm)}]
\draw[thick] (-2.5,0) -- (0.5,0)--(0.5,1);
\fill[white] (0.4,0.02) rectangle (0.6,0.08);
\draw[thick] (-1.5,1) -- (-1.5,0.05) -- (0.5,0.05) -- (1.15,0.05);
\draw[thick] (1.5,1) -- (1.5,0) -- (2.5,0);
\draw[thick, snake it] (-0.5,0.05) -- (-0.5,1);
\node at (-2.6,0) {$1$};
\node at (-1.5,1.2) {$2$};
\node at (-.5,1.2) {$3$};
\node at (.5,1.2) {$4$};
\node at (1.5,1.2) {$5$};
\node at (2.6,0) {$6$};
\node at (1.32,0.05) {$\partial$};
\node at (-1.4,-0.3) {${B}$};
\node at (-0.5,-0.3) {$\tilde{B}{B}$};
\node at (0.5,-0.3) {$\square \tilde{B}\tilde{Z}$};
\node at (1.5,-0.35) {$Z$};
\end{tikzpicture} 
 \, .
\end{align}
It is clear that the needed modification to the Lagrangian coming from these terms takes the following form:
 \begin{align}
\Delta  \mathcal{L}_6 \sim  \text{Tr} \, \Big( Z^\mu \Box \tilde{Z}_\mu + 
 [ \partial_\nu A^\mu ,  A_\mu ] Z^\nu 
+  [ A^\mu , \tilde{Z}^\nu ] \big( B_{\mu \nu} 
-   \Box \tilde{B}_{\mu \nu} \big) \Big) \, .
\end{align}
where we fixed the relative couplings by using the normalization freedom of the kinetic term, as well as the rescaling freedom $Z \rightarrow \alpha Z$ and $\tilde{Z} \rightarrow \tilde{Z} /\alpha$ that leaves the kinetic term invariant. There is only one free parameter, which we can take to be the overall normalization, and we find that it is equal to unity in order to match the correct $N^{(2)}$ contribution.

There are further mismatching monomials that we need to deal with.  The $N^{(2)}$ numerators contain problematic terms such as $\ee{2}{3} \, \ee{5}{6} \, \ez{1}{2} \, \varepsilon_4{\cdot}(p_6-p_5) \, \z_2^2$, where no Lorentz indices are crossing the central half-ladder propagator, and also the mass dimensions are unbalanced. This suggests that we need to introduce a pair of scalar fields $X$ and $\tilde{X}$ of mass dimension two and zero respectively. Again, we need to ensure that the new fields do not modify the five-point numerators.  Consider the new interactions $AA \tilde{X}$ and $A X \tilde{Z}$ which contribute to the mentioned monomial through the diagram
\begin{equation}
\ee{2}{3} \ee{5}{6} \ez{1}{2}  \varepsilon_4 {\cdot} (p_5 - p_6) \, \z_2^2
 \ \ \ \longrightarrow \ \ \ 
\begin{tikzpicture}[baseline={(0, +0.3cm)}]
\draw[thick] (-2.5,0) -- (-1.9,0) ;
\draw[thick] (1.5,1) -- (1.5,0) -- (2.5,0);
\draw[thick] (0.5,1) -- (0.5,0) -- (1.15,0);
\draw[thick,dotted] (-0.5,0) -- (0.5,0);
\draw[thick] (-0.5,1) -- (-0.5,0)--(-1.5,0)--(-1.5,1);
\node at (-2.6,0) {$1$};
\node at (-1.5,1.2) {$2$};
\node at (-.5,1.2) {$3$};
\node at (.5,1.2) {$4$};
\node at (1.5,1.2) {$5$};
\node at (2.6,0) {$6$};
\node at (-1.5,-0.45) {$ \partial A $};
\node at (-0.5,-0.4) {$ \Box A \tilde{X} $};
\node at (0.5,-0.4) {$X \tilde{Z}$};
\node at (1.5,-0.45) {$Z$};
\node at (1.3,0.05) {$\partial$};
\node at (-1.7,0.05) {$\partial$};
\end{tikzpicture}
\, .
\end{equation}
As mentioned, there are more derivatives on the left half of the diagram than the right half, thus the imbalance of the dimensions of the $X$ and $\tilde{X}$ fields.  

Additionally, we add $A \tilde{X} \tilde{Z}$ interaction to address the offending term
\begin{align}
 \ee{1}{2} \ee{5}{6} \varepsilon_3{\cdot}(p_1-p_2)\varepsilon_4{\cdot}(p_6-p_5) \z_2^2  
  \ \ \ \longrightarrow \ \ \ 
\begin{tikzpicture}[baseline={(0, +0.3cm)}]
\draw[thick] (-2.5,0) -- (-1.5,0) -- (-1.5,1) ;
\draw[thick] (1.5,1) -- (1.5,0) -- (2.5,0);
\draw[thick] (0.5,1) -- (0.5,0)  ;
\draw[thick,dotted] (-0.5,0) -- (0.5,0) ;
\draw[thick] (1.15,0) -- (0.5,0);
\draw[thick] (-0.5,1) -- (-0.5,0);
\draw[thick] (-0.5,0) -- (-1.15,0);
\node at (-2.6,0) {$1$};
\node at (-1.5,1.2) {$2$};
\node at (-.5,1.2) {$3$};
\node at (.5,1.2) {$4$};
\node at (1.5,1.2) {$5$};
\node at (2.6,0) {$6$};
\node at (-1.4,-0.45) {$Z$};
\node at (-0.5,-0.4) {$\square \tilde{Z} \scalarF$};
\node at (0.5,-0.4) {$\tilde{\scalarF}\tilde{Z}$};
\node at (1.5,-0.45) {$Z$};
\node at (-1.3,0.15) {$\overset{\leftarrow}{\partial}$};
\node at (1.3,0.05) {$\partial$};
\end{tikzpicture} \, .
\end{align}
We also need $\tilde{X}$ to interact with the $B$ field, to address offending terms of the form
\begin{equation}
\ee{2}{3} \, \ee{5}{6} \, \ez{1}{3} \, \varepsilon_4 {\cdot} (p_5 - p_6) \, \z_2^2
 \ \ \ \longrightarrow \ \ \ 
\begin{tikzpicture}[baseline={(0, +0.3cm)}]
\draw[thick] (-2.5,0) -- (-0.42,0) ;
\draw[thick] (1.5,1) -- (1.5,0) -- (2.5,0);
\draw[thick] (0.5,1) -- (0.5,0) -- (1.15,0);
\draw[thick,dotted] (-0.2,0) -- (0.5,0);
\draw[thick] (-0.5,1) -- (-0.5,0.05)--(-1.5,0.05)--(-1.5,1);
\node at (-2.6,0) {$1$};
\node at (-1.5,1.2) {$2$};
\node at (-.5,1.2) {$3$};
\node at (.5,1.2) {$4$};
\node at (1.5,1.2) {$5$};
\node at (2.6,0) {$6$};
\node at (-1.4,-0.4) {$\Box \tilde{B}$};
\node at (-0.5,-0.45) {$B \partial X$};
\node at (0.5,-0.4) {$\tilde{X} \tilde{Z}$};
\node at (1.5,-0.45) {$Z$};
\node at (-0.3,0.15) {$\overset{\leftarrow}{\partial}$};
\node at (1.3,0.05) {$\partial$};
\end{tikzpicture}
\, .
\end{equation}
The pair of scalar fields enjoys the same rescaling freedom as the auxiliary vectors, which we use to fix the coefficient of the $AA \tilde{X}$ interaction. The above three monomial structures are then enough to constrain all remaining coefficients of the needed interactions, which we find to be
 \begin{align}
\Delta  \mathcal{L}_6 &\sim \text{Tr} \, \Big( X \Box \tilde{X} + [ A^\mu,  \Box  A_\mu ] \tilde{X} -  [ A^\mu ,  X ] \tilde{Z}_\mu + \frac{1}{2}  [ A^\mu , \Box \tilde{X} ] \tilde{Z}_\mu - 2 [ A_\mu , B^{\mu \nu} ] \partial_\nu \tilde{X} \Big) \, .
\end{align}
This completes the construction of interactions that contribute to the six-point half-ladder numerator.

As a side remark that we will come back to later, note that one can rearrange the flow of Lorentz indices in the half-ladder diagrams using conservation of momentum, and it turns out that it is not strictly necessary to introduce a pair of scalar fields. Indeed, in the next subsection we show that it is possible to formulate a completion of the ${\cal L}_5$ Lagrangian in the NMHV sector using only the pair of vectors $Z$ and $\tilde{Z}$.

Finally, we need to address those diagrams that contribute to the non-half-ladder topology\footnote{Sometimes called star or Mercedes topology.} appearing at six points. Our Lagrangian does not yet get those contributions to match with what is predicted from the $N^{(2)}$ numerator. Assuming that no new fields are needed, the possible interactions are highly constrained by dimensional analysis and by the requirement that they must involve three auxiliary fields in order to not spoil the half-ladder numerators. Consider the following example of a missing term and corresponding diagram with a new $B\tilde{Z}\tilde{Z}$ interaction:
\begin{equation}
s_{12} \ee{3}{4} \ee{5}{6} \varepsilon_1 {\cdot} (p_5-p_6) \varepsilon_2 {\cdot} (p_3-p_4)
 \ \ \ \longrightarrow \ \ \ 
\label{eq:diagnh1}
\begin{tikzpicture}[baseline={(0, +0.3cm)}]
\draw[thick] (-2.5,0) -- (1.15,0);
\draw[thick] (1.5,0) -- (2.5,0);
\draw[thick] (-1.5,1) -- (-1.5,0.05)--(0,0.05)--(0,0.71);
\draw[thick] (1.5,1) -- (1.5,0) -- (2.5,0);
\draw[thick] (0.71,1.42) -- (0,1) -- (-0.71,1.42);
\node at (-2.6,0) {$1$};
\node at (-1.5,1.2) {$2$};
\node at (-.81,1.6) {$3$};
\node at (.81,1.6) {$4$};
\node at (1.5,1.2) {$5$};
\node at (2.6,0) {$6$};
\node at (-1.4,-0.4) {$\Box \tilde{B}$};
\node at (0,-0.4) {$B \tilde{Z}\tilde{Z}$};
\node at (1.5,-0.45) {$Z$};
\node at (-0.05,0.85) {\rotatebox[origin=c]{90}{$\partial$}};
\node at (1.3,0.05) {$\partial$};
\end{tikzpicture} \, . 
\end{equation}
We find that a suitable interaction has the form
\begin{equation}
  \text{Tr} \Big\{   [\tilde{Z}_{\mu} ,\tilde{Z}_{\nu}] \big( \beta  B^{\mu \nu}
+ (1-\beta)  \Box \tilde{B}^{\mu \nu} \big) \Big\}\,,
\end{equation}
but at this multiplicity we cannot yet fix the free parameter $\beta$. There is one more needed interaction which is rather simple, and fully constrained at six points, it is
\begin{equation}
 4  \text{Tr}   \Big\{ [B^{\mu \nu} ,\partial_\mu\tilde{B}_{\nu\rho} ] \tilde{Z}^{\rho} \Big\}\,. 
\end{equation}

This finishes the construction of the duality-satisfying six-point NMHV Lagrangian  $\mathcal{L}_6=\mathcal{L}_5+  \Delta  \mathcal{L}_6$, where all the new terms are assembled as
 \begin{align}
\Delta  \mathcal{L}_6 &= \text{Tr} \, \Big( Z^\mu \Box \tilde{Z}_\mu + X \Box \tilde{X}+ 
  [\partial_\nu A^\mu ,A_\mu]  Z^\nu +  [A^\mu, \Box  A_\mu ]\tilde{X}
+ [A_\mu, \tilde{Z}_\nu ]\big( B^{\mu \nu} 
-   \Box \tilde{B}^{\mu \nu} \big) \notag \\
& \quad \quad - [A^\mu, X ]\tilde{Z}_\mu + \frac{1}{2} [A^\mu, \Box \tilde{X}] \tilde{Z}_\mu - 2 [A_\mu, B^{\mu \nu} ] \partial_\nu \tilde{X} +  [\tilde{Z}_{\mu} ,\tilde{Z}_{\nu}] \big( \beta  B^{\mu \nu}
+ (1-\beta)  \Box \tilde{B}^{\mu \nu} \big) \notag \\
& \quad \quad  + 4  [B^{\mu \nu} ,\partial_\mu\tilde{B}_{\nu\rho} ] \tilde{Z}^{\rho} \Big) \, ,
\end{align}
and, as already mentioned, the $\beta$ parameter is not yet fixed.

Moving on to seven points, interactions of the form $\partial A Z \tilde{Z}$ and $\partial A X \tilde{X}$ can now partake in the half-ladder diagrams. These rather simple contributions are needed for generating correct  half-ladder factors $u_i$, which in hindsight is not surprising. 
Their coefficients in the Lagrangian can be fixed by computing the terms $\ee{1}{2} \ee{6}{7} \varepsilon_3 {\cdot} (p_1-p_2) \varepsilon_4 {\cdot} (p_6-p_7) u_5  \z_2^2$ and $\ee{1}{2} \ee{6}{7} \varepsilon_3 {\cdot} (p_1-p_2) \varepsilon_5 {\cdot} (p_6-p_7) u_4  \z_2^2$, and comparing the result to the predicted $N^{(2)}$ numerator. Furthermore, a new interaction $B \tilde{X} \tilde{Z}$ is required for the non-half-ladder graphs, and finally the unknown parameter $\beta$ from $\Delta  \mathcal{L}_6$ is now constrained to $\beta = 1/2$.    

Thus we conclude that the seven-point corrections to the duality-satisfying NMHV Lagrangian consists of the three terms
\begin{align}
\Delta \mathcal{L}_7 &= \text{Tr} \, \Big( \! -2  [A_\mu , Z_\nu] \partial^\mu \tilde{Z}^\nu -2  [A_\mu , \scalarF ] \partial^\mu \tilde{\scalarF}  - 2 [B^{\mu \nu}, \partial_\nu \tilde{X}] \tilde{Z}_\mu \Big) \, .
\end{align}
We find no further higher-multiplicity corrections, by explicitly computing and checking the properties of color-kinematics duality and gauge invariance for all NMHV numerators and amplitudes through ten points.
 
 Based on the robust patterns observed up to multiplicity ten, we conjecture that the following assembled NMHV Lagrangian computes all BCJ numerators and NMHV amplitudes to any multiplicity at tree level:
\begin{align}\label{eq:complete}
\mathcal{L} = \mathcal{L}_5 &+ \text{Tr} \, \Big( Z^\mu \Box \tilde{Z}_\mu + X \Box \tilde{X}+ 
  [\partial_\nu A^\mu, A_\mu]  Z^\nu +  [A^\mu, \Box  A_\mu] \tilde{X} -2  [A_\mu, Z_\nu ]\partial^\mu \tilde{Z}^\nu  \notag \\
  & \quad -2  [A_\mu, \scalarF ]\partial^\mu \tilde{\scalarF} + [A_\mu ,\tilde{Z}_\nu] \big( B^{\mu \nu} 
-   \Box \tilde{B}^{\mu \nu} \big)  - [A^\mu, X ]\tilde{Z}_\mu + \frac{1}{2} [A^\mu, \Box \tilde{X}] \tilde{Z}_\mu \notag \\
& \quad - 2 [A_\mu, B^{\mu \nu}] \partial_\nu \tilde{X}  +  \frac{1}{2} [\tilde{Z}_{\mu} ,\tilde{Z}_{\nu}] \big(   B^{\mu \nu}
+  \Box \tilde{B}^{\mu \nu} \big)  + 4  [B^{\mu \nu} , \partial_\mu\tilde{B}_{\nu\rho} ] \tilde{Z}^{\rho} \notag \\
& \quad - 2 [B^{\mu \nu} ,\partial_\nu \tilde{X}] \tilde{Z}_\mu \Big) \, . 
\end{align}
This is the simplest Lagrangian that we have found in this paper. A natural question to ask next is: how unique is it? 

\subsection{How unique is the NMHV Lagrangian?}
The way in which we obtained the NMHV Lagrangian \eqref{eq:complete} does not give strong evidence for its uniqueness. To check how unique it really is, we constructed a more general ansatz that is vastly larger in complexity compared to the above construction. Again, we constrained the results by checking color-kinematics duality and gauge invariance, but we were more meticulous in keeping track of our assumptions. This is useful for future work, where these assumptions can be further relaxed.

Our enlarged ansatz for the NMHV Lagrangian is subject to the following assumptions:
\begin{enumerate}
\item{The simple Lagrangian ${\cal L}_5$ \eqref{eq:l5} is still assumed to give the bi-scalar numerators.} 
\item{No additional fields beyond those in the previous sections are used. That is, only the  tensors $B$, $\tilde{B}$, vectors $Z$, $\tilde{Z}$ and scalars $X$, $\tilde{X}$ appear as auxiliary fields.}
\item{Kinetic terms do not mix fields: $\mathcal{L}_2 = \text{Tr} \, \Big( \frac{1}{2} A^\mu \Box A_\mu + B^{\mu \nu} \Box \tilde{B}_{\mu \nu} + Z^\mu \Box \tilde{Z}_\mu + X \Box \tilde{X} \Big) $.}
\item To preserve the four-point numerator, a pair of external $A$'s can either source $Z$ or $\tilde{Z}$. As before, we choose to exclude $AA \tilde{Z}$ interactions.
\item{Two-derivative interactions always appear as a d'Alembertian $\Box$. This makes it manifest that the interaction contributes at most to polarization-power two.}
\item{We make an {\it ad hoc} simplifying choice to exclude interactions $AZZ$ and $A \tilde{Z} \tilde{Z}$, in order the make the ansatz space more manageable.}
\end{enumerate}
After taking care of the rescaling freedom of the auxiliary fields, the Lagrangian ansatz we obtain with the above constraints has 174 free parameters. We constrain the numerators generated by the ansatz up to eight points by comparison with the predicted numerators $N^{(2)}$, as obtained from linear combinations of the bi-scalar numerators coming from ${\cal L}_5$. Note these constrains are non-linear equations in the free parameters of the Lagrangian ansatz, hence the equation system is non-trivial to deal with.   

With the constraints imposed up to eight points, this only fixes a subset of all parameters and we are left with 129 free coefficients. At nine points, we find that all numerator topologies generated by the Lagrangian ansatz are independent of the leftover coefficients. It is difficult to go to higher points due to the non-linearities, and we stop trying to find further constraints. Instead, we now seek solutions that minimize the number of terms in the Lagrangian. We find four solutions which all give 13-term Lagrangians (not counting kinetic terms or $\mathcal{L}_5$ interactions). Provided that $\partial \cdot A =0$, all solutions coincide (up to total derivatives) to give the same Lagrangian \eqref{eq:complete} we found in the previous section. Thus, it seems reasonable to think that this is the simplest NMHV Lagrangian, given the above list of assumptions. 

For example, another simple solution gives a 14-term Lagrangian. It is reachable by deforming the 13-term Lagrangian $\mathcal{L}$ \eqref{eq:complete} by the following interactions:
\begin{equation}
\label{eq:lag14term}
\mathcal{L} - \mathcal{L}_\text{14-term} = \text{Tr} \, \Big( [A^\mu, \Box A_\mu] \tilde{X} + 4 [A_\mu, \partial_\nu B^{\mu \nu}] \tilde{X} + [A^\mu, \tilde{X}] \Box \tilde{Z}_\mu \Big) \, .
\end{equation}
A more interesting scenario would arise if we managed to find a solution that makes use of fewer  auxiliary fields. We find no solutions that discards of the vectors $Z$ and $\tilde{Z}$, but interestingly, the scalars $X$ and $\tilde{X}$ do seem to not be strictly necessary for all solutions.  Consider replacing the above {\it ad hoc} assumption 6, by the following new assumption:
\begin{enumerate}
\item[6'.]{Exclude the scalar fields $X$, $\tilde{X}$, but now include all interactions $AZZ$ and $A \tilde{Z} \tilde{Z}$.}
\end{enumerate}
We start with a 95 parameter ansatz and constrain it by comparing to the predicted $N^{(2)}$ numerators up to eight points. At nine points, there is a single non-half-ladder graph depending on one free parameter. To fix this parameter, we use the unique half-ladder graphs generated by our ansatz and through Jacobi identities construct the graph containing the free parameter. At this stage, we have 50 free coefficients. Assuming they do not enter the numerators at higher multiplicity, we look for the smallest possible Lagrangian and find six solutions with 20 terms (not counting kinetic terms or $\mathcal{L}_5$ interactions). 
We choose to give one such solution, 
\begin{align}\label{eq:complete2}
\mathcal{L}'  = \mathcal{L}_5 &+ \text{Tr} \, \Big(  Z^\mu \Box \tilde{Z}_\mu - [A^\mu ,\partial_\nu A_\mu] Z^\nu + [A^\mu, Z^\nu ]\partial_\nu Z_\mu -\frac{3}{4} [A^\mu, Z_\mu] \partial \cdot Z - [A^\mu, \tilde{Z}_\mu] \partial \cdot \tilde{Z}   \notag \\
& \quad -2 [A^\mu, Z^\nu] \partial_\mu \tilde{Z}_\nu + \frac{1}{2} [A^\mu, \partial \cdot Z] \tilde{Z}_\mu -2 [A^\mu, \partial_\nu Z_\mu ]\tilde{Z}^\nu - \frac{3}{2} [A^\mu, Z_\mu] \partial \cdot \tilde{Z}  \notag \\
& \quad -\frac{3}{2} [A^\mu, B_{\mu \nu}] Z^\nu - [A^\mu, B_{\mu \nu}] \tilde{Z}^\nu - \frac{1}{2} [A^\mu, \Box \tilde{B}_{\mu \nu}] Z^\nu + [A^\mu, \Box \tilde{B}_{\mu \nu}] \tilde{Z}^\nu   \\
& \quad + \frac{9}{8} [B^{\mu \nu}, Z_\mu ]Z_\nu  + \frac{3}{2} [B^{\mu \nu} ,Z_\mu] \tilde{Z}_\nu + \frac{1}{2} [B^{\mu \nu}, \tilde{Z}_\mu ]\tilde{Z}_\nu + \frac{1}{8} [\Box \tilde{B}^{\mu \nu}, Z_\mu] Z_\nu \notag \\
& \quad - \frac{1}{2} [\Box \tilde{B}^{\mu \nu}, Z_\mu] \tilde{Z}_\nu  \notag + \frac{1}{2} [\Box \tilde{B}^{\mu \nu} ,\tilde{Z}_\mu ]\tilde{Z}_\nu -2 [B^{\mu \nu}, \partial_\mu \tilde{B}_{\nu \rho}] Z^\rho + 4 [B^{\mu \nu} ,\partial_\mu \tilde{B}_{\nu \rho} ]\tilde{Z}^\rho \Big) \, , \notag
\end{align}
as the other five solutions only differ from the one above by terms proportional to a divergence ($\partial \cdot A$, $\partial \cdot Z$ or $\partial \cdot \tilde{Z}$).
It would be interesting to more broadly explore the ansatz space of NMHV Lagrangians, with the above six assumptions further relaxed, but we leave it for future work.  

\subsection{Comments on the N${}^2$MHV sector}
The complete six-point BCJ numerator must contain also terms of polarization power three, which are needed for the N{}$^2$MHV amplitudes. Unfortunately, the Lagrangian ansatz space we have considered does not permit solutions which correctly reproduces this sector. This is not very surprising. Since, as a minimal extension of our ans\"{a}tze, we would need to introduce a three-form auxiliary field (or even rank-three tensors of mixed symmetries). This is clear from considering a six point BCJ numerator that necessarily~\cite{Chen:2021chy} contains terms of the form
\begin{equation}
\begin{tikzpicture}[baseline={(0, +0.3cm)}]
\draw[thick] (-2.5,-0.05) -- (2.5,-0.05);
\draw[thick] (-1.5,0) -- (1.5,0);
\draw[thick] (1.5,0) -- (1.5,1);
\draw[thick] (-1.5,0) -- (-1.5,1);
\draw[thick] (.5,0.05) -- (0.5,1);
\draw[thick] (-.5,0.05) -- (-.5,1);
\draw[thick] (-.5,0.05) -- (.5,0.05);
\node at (-2.6,-0.1) {$1$};
\node at (-1.5,1.2) {$2$};
\node at (-.5,1.2) {$3$};
\node at (.5,1.2) {$4$};
\node at (1.5,1.2) {$5$};
\node at (2.6,-0.1) {$6$};
\end{tikzpicture} 
     \ \ \sim\ \ 
     \varepsilon_1{\cdot} \varepsilon_6\, \varepsilon_2{\cdot} \varepsilon_5\, \varepsilon_3{\cdot}\varepsilon_4 \,\z_i^2 \, p_j{\cdot} p_k \, ,
\end{equation}
which in $D$-dimensions cannot be removed by generalized gauge transformations that maintain locality. The generalized gauge freedom of local crossing-symmetric $D$-dimensional six-point numerators were analysed in ref.~\cite{Chen:2021chy}. From the triple-solid line in the middle of the above diagram, it is clear that a rank-three auxiliary field is propagating. There could in principle also exist higher-rank fields since momentum dot products $p_j{\cdot} p_k$ may be contracted across this central propagator. Also, the type of tensors needed is unclear, it could be forms or tensors of mixed symmetries. We leave the problem of reproducing the $\text{N}^2$MHV and higher sectors to future work.

\section{Towards MHV numerators at one loop\label{sec:one-loop}}
Having obtained Lagrangians that generate BCJ numerators it is interesting to check if they can also generate BCJ numerators at loop level. Loop color-kinematics duality does not necessarily follow from tree level, since loops also contain off-shell internal states that are absent at tree level by the equations of motion. 

\subsection{Polarization power zero at one loop}
To obtain the one-loop polarization-power zero numerators (valid for all-plus and one-minus helicity YM sectors), we glue legs $1$ and $n$ of the tree-level half-ladder numerators, with appropriate contributions from Faddeev–Popov ghosts $c,\bar{c}$ to remove unphysical degrees of freedom,
\begin{equation}
    N^{(0)}_{\textrm{1-loop}}(1,2,\ldots,n) = N^{(1)}(\ell,1,2,\ldots,n,-\ell) + N(c,1,2,\ldots,n,\bar{c})+N(\bar{c},1,2,\ldots,n,{c}) \, .
\end{equation}
It should be understood that the states labeled by the loop momentum $\ell$ and $-\ell$ are contracted, and the two ghost numerators come with a minus sign after the sewing of the states since the $c$ and $\bar{c}$ fields are fermionic. 
As we have shown in section \ref{MHVsection}, the Feynman rules for the bi-scalar sector automatically obey the color-kinematics duality so long as the external vector fields are transverse. This assumption holds for external states as well as subdiagrams that are of polarization power zero, just as argued below \eqn{diffeo_current}. 

All we have to do is add the ghosts in such a way that they respect the color-kinematics duality, and this is achieved by the standard Faddeev-Popov ghost Lagrangian
\begin{eqnarray}
\mathcal{L}_\text{ghost}^{(0)} &=& \text{Tr} \, \Big(
c\Box \bar{c} +A^\mu c\partial_\mu \bar{c} \Big) \, .
\end{eqnarray}
The interaction term here is identical to the effective interactions of the bi-scalar sector so color-kinematics duality works out automatically when external states are transverse. Since the diagrams that contribute here come from standard YM Feynman rules, the auxiliary fields are not yet present. 

Let us give the duality-satisfying $n$-gon master numerators that are valid for the all-plus and one-minus helicity one-loop amplitudes (see also refs.~\cite{Boels:2013bi,Edison:2022jln}),
%
\begin{equation}
    N^{(0)}_{\textrm{1-loop}}(1,\ldots,n) = \textrm{Tr}(W_1\cdots W_n)+ \textrm{Tr}(\widetilde W_1\cdots \widetilde W_n) - (2+D)U_{1,n}\, ,
\end{equation}
where $D=\textrm{Tr}(1)$ is the dimension, $U_{1,n}$ is defined in \eqn{Udef} and the matrices are
\begin{equation}
    (W_i)^{\mu\nu} = u_i \eta^{\mu\nu}+ 
    2 \varepsilon_i^\mu p_i^\nu 
    \,,~~~~
    (\widetilde W_i)^{\mu\nu} = u_i \eta^{\mu\nu}- 
    2 p_i^\mu \varepsilon_i^\nu 
    \, .
\end{equation}
The region momenta present in $u_i=2\varepsilon_i{\cdot} x_i$ are here defined to include the loop momentum $x_i = \ell+ \sum_{j=1}^{i}p_j$.
The above numerators can be checked to give the correct maximal~\cite{Edison:2022jln} and non-maximal cuts, since by construction they make use of the standard YM three-point vertex and there are no contact terms present in the amplitudes.

\subsection{Comment on one-loop MHV numerators}
As already emphasized, tree-level all-multiplicity color-kinematics duality is in general insufficient to infer that the loop-level duality holds. Indeed, we find obstructions in realizing one-loop BCJ numerators in the MHV sector from the Feynman rules derived in previous sections. 
At polarization-power one, in particular, the numerators receive contributions from the new fields we added to the Lagrangian. As before, we have contributions from the tree-level polarization-power one sector after gluing half-ladder numerators,
\begin{equation}
    N^{(1)}_{\textrm{1-loop}}(1,2,\ldots,n) \supset N^{(1)}(\ell,1,2,\ldots,n,-\ell) \, ,
\end{equation}
but now we must add the glued half-ladder numerator at polarization power two as well,
\begin{equation}
    N^{(1)}_{\textrm{1-loop}}(1,2,\ldots,n) \supset N^{(2)}(\ell,1,2,\ldots,n,-\ell) \, .
\end{equation}
In addition, to not break crossing symmetry we should allow for the $B$, $Z$ and $X$ fields to cross the sewn loop line. This makes it possible for the auxiliary fields to propagate all the way around the loop, which is not expected to give reasonable contributions. Thus, we may attempt to project out certain contributions by adding new ghosts $\{b, z, \chi \}$ for the auxiliary fields, giving the ghost Lagrangian 
\begin{eqnarray} \label{GL1}
\mathcal{L}_\text{ghost}^{(1)} &=&
\text{Tr} \, \Big(
c \Box \tilde{c} + b^{\mu\nu}\Box \tilde{b}_{\mu\nu} + 
A^\mu c\partial_\mu \bar{c} + 
4  \partial_\nu \tilde{b}_{\mu \rho} [A^\mu ,b^{\nu \rho}] - 2 \partial_\rho \tilde{b}_{\mu \nu} [ A^\rho,  b^{\mu \nu}] \nonumber \\
&&\quad\quad +z\Box \tilde{z} + \chi\Box\tilde{\chi} -2  [A_\mu, z_\nu ]\partial^\mu \tilde{z}^\nu -2  [A_\mu, \chi ]\partial^\mu \tilde{\chi}
\Big) \, .
\end{eqnarray}
As usual, the ghost fields cannot be sourced and so they only contribute through propagating in a complete closed loop. Note that these ghosts are added by hand in order to cancel closed loops of unphysical fields, and not derived from some underlying gauge-fixing procedure.

We performed a few crude tests to the one-loop numerators as obtained from our Lagrangians, and found that for the 13-term Lagrangian in \eqn{eq:complete} already the three-point one-loop numerators are not well behaved, meaning they do not give gauge-invariant unitarity cuts. For the 14-term Lagrangian obtained by the deformation \eqref{eq:lag14term}, the maximal cuts work as tested up to four points, but the next-to maximal cuts do not. For example, the following contribution to the box diagram appears to not be correct:
\begin{eqnarray}
\begin{tikzpicture}[baseline={(0, -0.6cm)}]
\draw[thick] (0,0) -- (1,0);
\draw[thick] (1,0) -- (1,-1);
\draw[thick] (1,-1) -- (0,-1);
\draw[thick] (0,-1) -- (0,0);
\draw[thick,snake it] (-0.5,0.5) -- (0,0);
\node at (-0.7,0.7) {$1$};
\draw[thick,snake it] (1.5,0.5) -- (1,0);
\node at (1.7,0.7) {$2$};
\draw[thick,snake it] (1.5,-1.5) -- (1,-1);
\node at (1.7,-1.7) {$3$};
\draw[thick,snake it] (-0.5,-1.5) -- (0,-1);
\node at (-0.7,-1.7) {$4$};
\node at (0.1,-0.5) {$\uparrow$};
\node at (0.2,0.2) {${}_{\square\tilde{B}}$};
\node at (0.9,0.17) {${}_{B}$};
\node at (1.2,-0.17) {${}_{\scalarF}$};
\node at (1.2,-0.8) {${}_{\tilde{\scalarF}}$};
\node at (0.8,-1.2) {${}_{\tilde{Z}}$};
\node at (0.2,-1.2) {${}_{Z}$};
\node at (-0.2,-0.8) {${}_{A}$};
\node at (-0.2,-0.2) {${}_{A}$};
\end{tikzpicture} 
\ \ \sim \ \ 
(\varepsilon_3\cdot \z_3) \z_1^2 \varepsilon_4^{[\mu}\varepsilon_1^{\nu]}(2p_{2\mu} + \z_{2\mu})\varepsilon_{2\nu} \, .
\end{eqnarray}
Specifically, the term that spoils gauge invariance is proportional to $\varepsilon_1\cdot \varepsilon_2$.
It is clear that the ghost Lagrangian \eqref{GL1} does not remove this diagram, nor is it removed by introducing further obvious vector-ghost interactions, inspired by the fields already present. While it should be possible to introduce fine-tuned ghosts and interactions that precisely cancel this diagram, it is a non-trivial task to ensure the new terms are consistent with color-kinematics duality and gauge invariance for all higher-multiplicity MHV numerators. We leave the problem of formulating duality-satisfying one-loop-compatible Lagrangians to future work.

\section{Conclusions}
\label{sec:conclucions}
In this paper, we considered the problem of constructing Lagrangians that manifest color-kinematics duality for YM theory. Such explicit Lagrangians can be used to compute BCJ numerators, as well as give non-trivial clues to the mathematical structure underlying color-kinematics duality. While duality-satisfying tree-level numerators are known to any multiplicity, finding corresponding Lagrangian descriptions appears to be more challenging. The problem simplifies by restricting to helicity sectors of YM, and in this paper we fully address the NMHV sector.      

As a first step, we found a simple Lagrangian~\eqref{eq:l5} that is fully equivalent to the standard YM Lagrangian at tree level, and it computes BCJ numerators in the bi-scalar subsector of the NMHV sector. This Lagrangian was constructed by first resolving the four-gluon contact term using a pair of auxiliary two-forms fields, and subsequently deforming with new cubic interactions involving these auxiliary fields. The bi-scalar numerators can be computed to any multiplicity, and they provide complete information via \eqn{ym_n_from_scalar} to obtain all tree-level NMHV numerators. These provide a clear target for what a complete NMHV Lagrangian should reproduce.  

Next we searched for a complete NMHV Lagrangian, and we found that by introducing at most two additional pairs of auxiliary fields, of vector and scalar type, there are several solutions for such Lagrangians. Using a larger ansatz, we found that solutions also exists if the scalar is removed at the cost of additional interactions between the remaining fields. Because of the large freedoms of the Lagrangians, we chose to present the simplest solutions, given in eqs.~\eqref{eq:complete} and \eqref{eq:complete2}, which were explicitly tested through ten points and conjectured to work to all multiplicities at tree level. With our current limited Lagrangian ansatz space, we cannot obtain the N${}^2$MHV and higher sector contributions to the BCJ numerators. Nor are the presented duality-satisfying NMHV Lagrangians equivalent to YM in the N${}^2$MHV sector and beyond, unlike the simple Lagrangian~\eqref{eq:l5} first found. It would be desirable to revisit this problem in the future, and repeat the ansatz construction of duality-satisfying NMHV Lagrangian while maintaining gauge covariance at intermediate steps such that the N${}^2$MHV and higher sectors are not spoiled, even if they might not fully enjoy color-kinematics duality.  

We briefly discussed the need of higher-rank tensor auxiliary fields to reproduce N${}^2$MHV sector BCJ numerators. It is clear that in a covariant ($D$-dimensional) and local formalism it is unavoidable to encounter, at the very minimum, three-form fields in a duality satisfying N${}^2$MHV Lagrangian. However, the minimal set of needed auxiliary fields in the N${}^2$MHV sector is something that needs further studies. Currently the main challenge in brute-force constructions of duality-satisfying Lagrangians is to predict the number and types of needed auxiliary fields. Clearly it would be desirable to better understand the general structure and need of such fields such that more refined Lagrangian ans\"{a}tze can be constructed. Small finely tuned ans\"{a}tze would speed up progress because the equation systems encountered are non-linear in the ansatz parameters, and the solutions contains large redundancy making them difficult to analyze.

Even in the NMHV sector there are more questions to be answered. Can one remove some of the assumptions that went into our constructions and perhaps obtain much simpler Lagrangians? For example, hints from Chern-Simons-type Lagrangians~\cite{Ben-Shahar:2021doh,Ben-Shahar:2021zww} suggest that it should be beneficial to look for more intricate kinetic terms than the diagonal ones used in this paper. Furthermore, gauge covariance plays no role in the current construction and this is likely an oversight that should be addressed in more refined attempts. While our NMHV Lagrangians fit on a few lines, it is fair to say that their complexity is likely artificially high compared to more optimal duality-satisfying rewritings of the YM Lagrangian that might be found in the future. Nevertheless, we have taken critical steps in this research program by finding the first examples of NMHV Lagrangians that: 1) use very few auxiliary fields 2), have very simple structure in the bi-scalar subsector, and 3) give local all-multiplicity BCJ numerators.

\bigskip

\section*{Acknowledgments}

We thank Zvi Bern, Lucile Cangemi, Gang Chen, Paolo Pichini, Oliver Schlotterer, Fei Teng, Tianheng Wang and Maxim Zabzine for enlightening discussions related to this work. This research was supported in part by the Knut and Alice Wallenberg Foundation under grants KAW 2018.0116 ({\it From Scattering Amplitudes to Gravitational Waves}) and KAW 2018.0162 ({\it Exploring a Web of Gravitational Theories through Gauge-Theory Methods}), as well as the Ragnar S\"{o}derberg Foundation (Swedish Foundations’ Starting Grant).

\bibliographystyle{JHEP}
\bibliography{bib}

\end{document}